\documentclass[a4,12pt,draftclsnofoot,onecolumn]{IEEEtran}

\usepackage[T1]{fontenc}% optional T1 font encoding

\usepackage[none]{hyphenat}
\usepackage[utf8]{inputenc}
\usepackage[english]{babel}
\usepackage{subcaption}
\usepackage{tikz}
\usetikzlibrary{calc}
\usetikzlibrary{arrows.meta}
\tikzset{>=latex}

\usepackage{verbatim}
\usepackage{graphicx}
\usepackage{float}
\usepackage{hyperref}
\usepackage{lipsum}
\usepackage{stfloats}
\usepackage{amsmath,amsthm}
\usepackage{amssymb}
\usepackage{mathtools}
\usepackage{epstopdf}
\usepackage{booktabs,dcolumn,caption}
\usepackage{multirow}
\usepackage{caption}
\usepackage{subcaption}
\usepackage{ragged2e}
\usepackage{cite}
\usepackage{censor}
\usepackage{array}
\newcolumntype{C}[1]{>{\centering\arraybackslash}p{#1}}
\newcolumntype{L}[1]{>{\RaggedRight\arraybackslash}p{#1}}
\usepackage{wasysym}
\newcommand*\diff{\mathop{}\!\mathrm{d}}

\usepackage{mathrsfs}
\usepackage{xcolor}
\usepackage{colortbl}
\usepackage{color,soul}
\usepackage{caption,hypcap}
\usepackage{cleveref}
\usepackage{suffix}

\usepackage{cleveref}
\usepackage{scalerel}
\usepackage{tikz}
\usetikzlibrary{svg.path}

\definecolor{orcidlogocol}{HTML}{A6CE39}
\tikzset{
  orcidlogo/.pic={
    \fill[orcidlogocol] svg{M256,128c0,70.7-57.3,128-128,128C57.3,256,0,198.7,0,128C0,57.3,57.3,0,128,0C198.7,0,256,57.3,256,128z};
    \fill[white] svg{M86.3,186.2H70.9V79.1h15.4v48.4V186.2z}
                 svg{M108.9,79.1h41.6c39.6,0,57,28.3,57,53.6c0,27.5-21.5,53.6-56.8,53.6h-41.8V79.1z M124.3,172.4h24.5c34.9,0,42.9-26.5,42.9-39.7c0-21.5-13.7-39.7-43.7-39.7h-23.7V172.4z}
                 svg{M88.7,56.8c0,5.5-4.5,10.1-10.1,10.1c-5.6,0-10.1-4.6-10.1-10.1c0-5.6,4.5-10.1,10.1-10.1C84.2,46.7,88.7,51.3,88.7,56.8z};
  }
}
\tikzset{near start abs/.style={xshift=0.1cm}}

\newcommand\orcidicon[1]{\href{https://orcid.org/#1}{\mbox{\scalerel*{
\begin{tikzpicture}[yscale=-1,transform shape]
\pic{orcidlogo};
\end{tikzpicture}
}{|}}}}

\DeclarePairedDelimiterX\MeijerM[3]{\lparen}{\rparen}%
{#3\delimsize\vert\,\begin{matrix}#1 \\ #2\end{matrix}}
\newcommand\MeijerG[8][]{%
	G^{\,#2,#3}_{#4,#5}\MeijerM[#1]{#6}{#7}{#8}}

\WithSuffix\newcommand\MeijerG*[7]{%
	G^{\,#1,#2}_{#3,#4}\MeijerM*{#5}{#6}{#7}}
\hypersetup{linkcolor=black,citecolor=black}

\begin{document}

\title{An Alternative Statistical Characterization of TWDP Fading Model}

\author{Almir~Maric%\textsuperscript{\orcidicon{0000-0001-5912-2967}}
,~\IEEEmembership{Member,~IEEE,}
        Pamela~Njemcevic%\textsuperscript{\orcidicon{0000-0002-3005-3934}}
        ,~\IEEEmembership{Member,~IEEE,}
        and~Enio~Kaljic%\textsuperscript{\orcidicon{0000-0003-1902-2608}}
        ,~\IEEEmembership{Member,~IEEE}
        %and~Vladimir~Lipovac,~\IEEEmembership{Member,~IEEE}
%\thanks{Manuscript received XXX; revised XXX and XXX; accepted XXX. This work was in partial supported by XXX. The associate editor coordinating the review of this paper and approving it for publication was XXX (\it{Corresponding author: Almir Maric}).}
\thanks{A. Maric, P. Njemcevic, and E. Kaljic are with the Department
of Telecommunications, Faculty of Electrical Engineering, University of Sarajevo, Sarajevo, Bosnia and Herzegovina.}
%\thanks{V. Lipovac is with the Department of Electrical Engineering and Computing, University of Dubrovnik, Dubrovnik, Croatia.}
\thanks{e-mail: almir.maric@etf.unsa.ba;}
}

\maketitle

\begin{abstract}
Two-wave with diffuse power (TWDP) is one of the most promising models for description of small-scale fading effects in emerging wireless networks. However, its current statistical characterization has several fundamental issues. Primarily, conventional TWDP parameterization is not in accordance with the model's underlying physical mechanisms. In addition, available TWDP expressions for PDF, CDF, and MGF are given either in integral or approximate forms, or as mathematically untractable closed-form expressions.
Consequently, the existing TWDP statistical characterization does not allow accurate evaluation of system performance (such as error and outage probability) in all fading conditions for most modulation and diversity techniques.
In this paper, the existing statistical characterization of the TWDP fading model is improved by overcoming some of the noticed issues. In this regard, physically justified TWDP parameterization is proposed and used for further calculations. Additionally, exact infinite-series PDF and CDF are introduced. Based on these expressions, the exact MGF of the SNR is derived in form suitable for mathematical manipulations.
The applicability of the proposed MGF for derivation of the exact average symbol error probability (ASEP) is demonstrated with the example of M-ary PSK modulation. Therefore, in this paper, M-ary PSK ASEP is derived as an explicit expression for the first time in the literature.
The derived expression is further simplified for large SNR values in order to obtain a closed-form asymptotic ASEP, which is shown to be applicable for \mbox{SNR > 20 dB}. All proposed expressions are verified by Monte Carlo simulation in a variety of TWDP fading conditions.
%%%%%%%SKRACENA VERZIJA
%Two wave with diffuse power (TWDP) is one of the most promising small-scale models. However, its statistical characterization has several  issues. Primarily, conventional TWDP parameterization is not in accordance with model's underlying physical mechanisms. Besides, available TWDP PDF and MGF are given either in integral or approximate forms, or as a mathematically untractable closed-form expressions.
%Consequently, the existing TWDP statistical characterization doesn't allow accurate evaluation of the effect of TWDP fading on system performances (such as error and outage probability) in all fading conditions.
%In this paper, the existing statistical characterization of TWDP fading model is improved. In that sense, physically justified TWDP parameterization is proposed and used for further calculations. Additionally, the exact infinite-series PDF and CDF are introduced. Based on these expressions, the exact SNR MGF  is derived, in suitable form for mathematical manipulations.
%The applicability of proposed MGF for derivation of the exact average symbol error probability (ASEP) is demonstrated on the example of M-ary PSK modulation. Derived expression is further simplified for large SNR values in order to obtain closed-form asymptotic ASEP, applicable for SNR > 20 dB. All proposed expressions are verified by Monte Carlo simulation, in varieties of TWDP fading conditions.

\end{abstract}

\begin{IEEEkeywords}
TWDP fading channel, MGF, M-ary PSK, ASEP.
\end{IEEEkeywords}

\IEEEpeerreviewmaketitle

\section{Introduction}
\label{sec:I}
\IEEEPARstart{D}{ue} to the tremendous growth of Internet data traffic, bandwidth requirements have become especially pronounced. To cope with these requirements, the fifth generation (5G) mobile network is emerging as the latest wireless communication standard. At the heart of this technology lies the use of millimeter wave (mmWave) frequency band.
%Wireless communication saw rapid development due to high bandwidth demand fuelled by growth of Internet data services. Now, the fifth generation (5G) is emerging as the latest mobile communication standard. At the heart of this technology lies the use of millimeter wave (mmWave) frequency bands. 
However, a signal propagating in mmWave band exhibits unique propagation properties, making traditional small-scale fading models inadequate and thus demanding more generalized  models. To address this issue, Durgin \textit{et al.}~\cite{Dur02} proposed the two-wave with diffuse power (TWDP) model, which assumes that the complex envelope consists of two strong specular components and many weak diffuse components. As such, it encompasses Rayleigh, Rician, and two-ray fading models as its special cases~\cite{Dur02}, simultaneously enabling modeling of both worse-than-Rayleigh and Rician-like fading conditions.

%\hl{Due to its applicability for modeling varieties of fading conditions,} 
In the last twenty years the TWDP model has been extensively studied theoretically (there are more than 5000 results on Google). Additionally, its existence is supported by practical evidences both in mmWave communication systems equipped with directional antennas or arrays~\cite{Rap15} and in wireless sensor networks deployed in cavity environments~\cite{Fro07}.
However, to the best of the authors' knowledge, there are at least two factors that motivate further studies of TWDP fading and its performance: 
\begin{enumerate}
    \item existing TWDP parameterization is not in accordance with the model's underlying physical mechanisms,
    %One of two parameters proposed by Durgin~\textit{et al.}~\cite{Dur02} is not in accordance with model's underlying physical mechanisms.
    \item analytical forms of the existing expressions for PDF and MGF disallow accurate evaluation of the effects of TWDP fading on system performance.
    %disable  error probability performances in all fading conditions implied by TWDP model, for most modulation and diversity schemes.  
    %\hl{The second one is related to a limited accuracy of the existing explicit envelope PDF. It is available only as an approximate expression, thus preventing accurate error probability evaluation in diverse TWDP fading conditions.} 
\end{enumerate}

To describe TWDP fading, Durgin et al.~\cite{Dur02} proposed two parameters, $K\geq 0$ and $0\leq\Delta \leq 1$, which reflect the relationship between specular and diffuse components and between the specular components themselves. However, it is striking that for a significant range of values ($0\leq\Delta\leq0.5$) the impact of $\Delta$ on the system performance metrics (e.g. ASEP and outage probability) is negligible (in fact, in some cases the corresponding curves almost overlap). This is obviously counterintuitive considering the physical meaning attributed to the parameter~$\Delta$. It is thus essential to examine this problem in depth. 

Regarding TWDP PDF expressions, they exist in integral~\cite[eq. (29)]{Dur02}, ~\cite[eq. (32)]{Dur02},~\cite[eq. (16)]{Rao14}, and approximate~\cite[eq. (17)]{Dur02} forms. Therefore, the exact evaluation of system performance metrics based on the existing integral expressions is not mathematically tractable, disabling direct observation of TWDP fading effects on system performance. 
Accordingly, the closed-form results of performance evaluation (e.g. error and outage probability, ect.) obtained from PDF are available 
only in approximate forms~\cite{Oh05,Sur08,Dix13,Sin15,Tan11,Hag11,Lee07,Hag12,Oh07, Sub14, Lu11, Lee07-1}. However, it has been shown that analysis based on an approximate PDF expression is accurate only for a narrow range of $K$ and $\Delta$ values~\cite{Kim17,Dur02}, which can only be used for description of limited fading conditions.  

To overcome these limitations, Rao~\textit{et al.}~\cite{Rao15} proposed an alternative approach to statistical characterization of TWDP fading based on the observation that the TWDP fading model can be expressed in terms of a conditional underlying Rician distribution. Thus, by invoking the observed similarities and the existing expressions of Rician fading, Rao~\textit{et al.} derived a novel form of TWDP MGF expression~\cite[eq. (25)]{Rao15}. Thereby, in contrast to previously derived approximate MGF expressions~\cite[eq. (8)]{Oh07}\cite[eq. (12)]{Dix13}, the one proposed in~\cite{Rao15} is given as a simple closed-form solution. However, this form is also not suitable for mathematical manipulations, and consequently, for calculation of the exact ASEP expressions for most modulation and diversity schemes. The exception is ABEP expressions for DBPSK  modulations as derived in~\cite{Rao14}. Accordingly, in order to accurately evaluate the effects of TWDP fading on ASEP, outage probability, etc., it is of tremendous importance to provide mathematically tractable PDF and MGF expressions.

Considering the above, our contributions are as follows: 
\begin{enumerate}
    \item We proposed alternative TWDP parameterization which is in accordance with model's underlying physical mechanisms.
    \item We introduced the exact convergent infinite-series TWDP envelope PDF and CDF expressions (previously derived in~\cite{Kos78, Kos96}). 
    %It is shown that proposed PDF has no limitations regarding TWDP parameters' values. 
    \item We derived an alternative exact form of SNR MGF based on the adopted CDF expression and proposed parameterization, which is shown to be suitable for mathematical manipulations. 
    
    \item Based on the obtained MGF, we derived M-ary PSK ASEP in exact infinite-series form, which is, to the best of our knowledge, the first such expression proposed to date.  %Derived expression is verified by Monte Carlo simulation.
    %By comparing results obtained using derived ASEP expression to those obtained by Monte Carlo simulation, it is shown that proposed analytical ASEP can be used for accurate evaluation of error rate performance of M-ary PSK receiver for diverse TWDP fading conditions. 
    \item We also derived asymptotic M-ary PSK ASEP as a simple closed-form expression, which tightly follows the exact one for the practical range of SNR values, i.e. for SNR~>~20~dB.
 \end{enumerate}
 
 The rest of the paper is structured as follows. In Section~\ref{sec:II}, the TWDP fading model is introduced and statistically described using alternative envelope PDF and CDF expressions, given in terms of newly proposed parameters. The alternative MGF of the SNR expression is derived in Section~\ref{sec:III}. 
 In Section~\ref{sec:IV}, the applicability of the proposed MGF for accurate performance analysis is demonstrated by deriving the exact and asymptotic M-ary PSK ASEP expressions, which are then verified by Monte Carlo simulation. The main conclusions are outlined in Section~\ref{sec:V}.
 
\section{TWDP fading model}
\label{sec:II}
In the slow, frequency nonselective fading channel with TWDP statistic, the complex envelope $r(t)$ is composed of two strong specular components $v_1(t)$ and $v_2(t)$ and many low-power diffuse components treated as a random process $n(t)$:
\begin{equation}
    \begin{split}
    \label{eq_1010}
        r(t) &= v_1(t) + v_2(t) + n(t) \\
        & = V_1\exp{\left(j\Phi_1\right)} + V_2\exp{\left(j\Phi_2\right)} + n(t)
    \end{split}
\end{equation}
Specular components are assumed to have constant magnitudes $V_1$ and $V_2$ and uniformly distributed phases $\Phi_1$ and $\Phi_2$ in $[0, 2\pi)$, while diffuse components are treated as a complex zero-mean Gaussian random process $n(t)$ with average power $2\sigma^2$. Consequently, the average power of a signal $r(t)$ is equal to $\Omega = V_1^2 + V_2^2 + 2\sigma^2$. 

\subsection{The revision of parameter \texorpdfstring{$\Delta$}{Δ}}
Conventional parameterization of TWDP fading, originally proposed in~\cite{Dur02}, introduced two parameters:
%In order to classify the shape of TWDP PDF, two parameters, $K$ and $\Delta$, are proposed in~\cite{Dur02} and defined as:  
\begin{equation}
\label{eq_2000}
   K \triangleq \frac{\text{average~specular~power}}{\text{diffuse~power}} = \frac{V_1^2+V_2^2}{2 \sigma^2}  
\end{equation}
and
\begin{equation}
    \label{eq_2002}
    \Delta \triangleq \frac{\text{peak~specular~power}}{\text{average~specular~power}} - 1 = \frac{2V_1 V_2}{V_1^2+V_2^2}
\end{equation}
Parameter $K$, $(0 \leq K < \infty)$, like in the Rician fading model, characterizes TWDP fading severity. Parameter $\Delta$, \mbox{$(0\leq \Delta\leq 1)$} for $V_1\geq0$, $V_2\geq0$, and $V_2 \leq V_1$, implicitly characterizes the relationship between the magnitudes of specular components. However, the physical justification of the relationship between $V_1$ and $V_2$, introduced by the definition of parameter $\Delta$ in (\ref{eq_2002}), is questionable. Namely, according to~\cite{Kos20} "\textit{for $0<\Delta<1$ there is a nonlinear relation between the magnitude of the specular components $V_1$ and $V_2$, i.e., $V_2=V_1 (1-\sqrt{1-\Delta^2})/\Delta$. However, the physical facts suggest a different conclusion about the relation between $V_1$ and $V_2$. In particular, according to the model for TWDP fading, specular components are constant and they are a consequence of specific propagation conditions. Since electromagnetic wave is propagating in a linear medium, a natural choice to appropriately characterize the relation between magnitudes $V_1$ and $V_2$ is given by $\Gamma\triangleq\ V_2/V_1$, where $V_2\leq V_1$. Seemingly,  parameters $\Delta$ and $\Gamma$ are both motivated by physical arguments. However, they do not have the same level of physical intuition.}".

Based on the above citation, it is necessary to further investigate the impact of nonlinear $\Delta$-based parameterization of TWDP statistics. Accordingly, parameters $K$ and $\Delta$ are written in terms of $V_2/V_1$, as:
\begin{equation}
    \label{eq_2009}
    \begin{split}
        K& =  \frac{V_1^2+V_2^2}{2\sigma^2}\\
        & = \frac{V_1^2}{2 \sigma^2}\left[1+\left(\frac{V_2}{V_1}\right)^2\right] = K_{Rice}(1 + \Gamma^2)
    \end{split}
\end{equation}
and 
\begin{equation}
    \label{eq_2007}
    \Delta = \frac{2 \frac{V_2}{V_1}}{1+\left(\frac{V_2}{V_1}\right)^2}
\end{equation}
where $\Gamma = V_2/V_1$ and $K_{Rice} = V_1^2/(2\sigma^2)$ represents the Rician parameter $K$ of a dominant specular component. 
Based on the above, parameter $K$ is also expressed in terms of $\Delta$, as:
\begin{equation}
    \begin{split}
    \label{eq_2011}
    K = & \frac{V_1^2+V_2^2}{2\sigma^2}\\    & =\frac{1}{2\sigma^2} \frac{2V_1V_2}{\Delta} \frac{V_1}{V_1}
    = \frac{V_1^2}{2 \sigma^2} \frac{2}{\Delta} \frac{V_2}{V_1} \\
    & = K_{Rice}2\frac{1-\sqrt{1-\Delta^2}}{\Delta^2}
    \end{split}    
\end{equation}
%and (\ref{eq_2009}) - (\ref{eq_2011}) are illustrated on Fig.~\ref{Figure01} - Fig.~\ref{Fig.2b}.

\begin{figure}[H]
    \centering
    \begin{minipage}[b,trim={0.15cm, 0.15cm, 0.15cm, 0.95cm},clip]{0.50\linewidth}
        \centering
        \includegraphics[trim={0.15cm, 0.15cm, 0.15cm, 0.95cm},clip,width=0.99\linewidth]{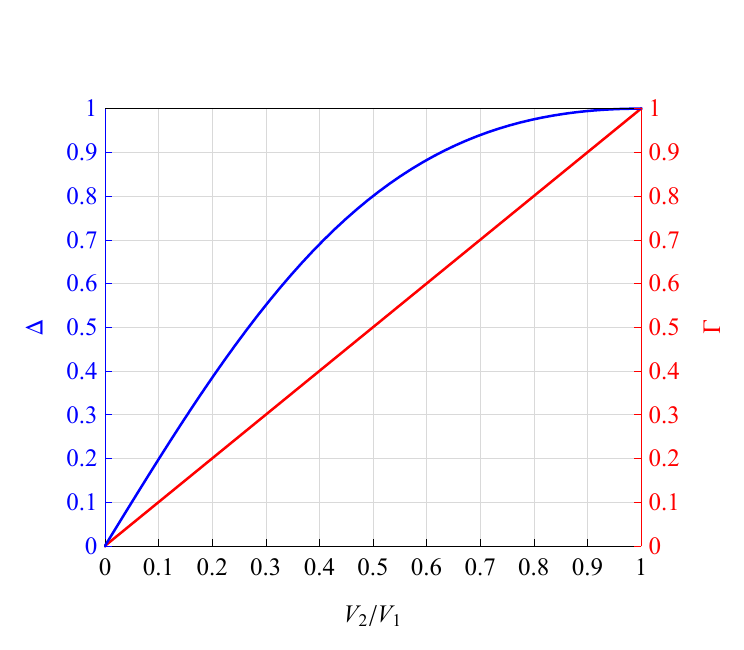}
        \caption{Dependence of $\Delta$ and $\Gamma$ on $V_2/V_1$}
        \label{Figure01}
        \vspace{4ex}
    \end{minipage}%%
%\end{figure}
%\begin{figure}[H]
    \vspace{0.051cm}
    \begin{minipage}[b,trim={0.15cm, 0.15cm, 0.15cm, 0.95cm},clip]{0.50\linewidth}
        \centering
        \includegraphics[trim={0.15cm, 0.15cm, 0.15cm, 0.95cm},clip,width=0.99\linewidth]{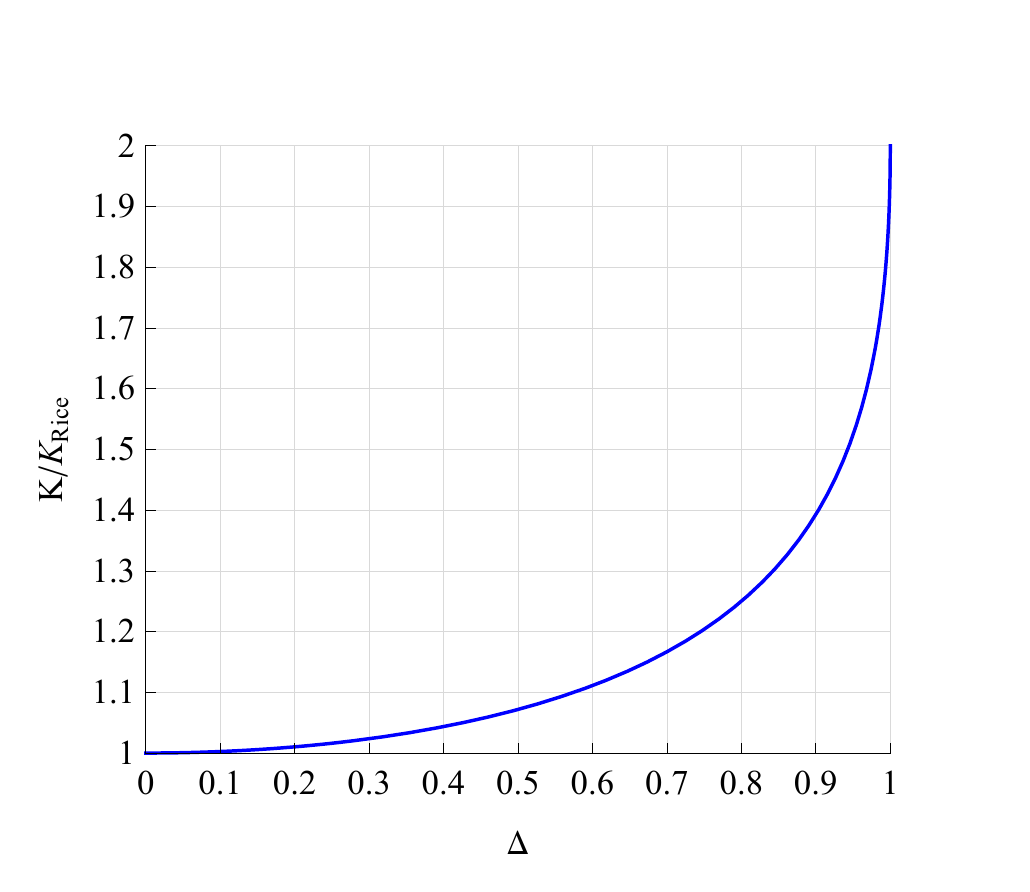}
        \caption{Dependence of $K/K_{Rice}$ on $\Delta$}
        \label{Fig.2a}
        \vspace{4ex}
    \end{minipage}%%
%\end{figure}
%\begin{figure}[H]
    \vspace{0.051cm}
    \begin{minipage}[b,trim={0.15cm, 0.15cm, 0.15cm, 0.95cm},clip]{0.50\linewidth}
        \centering
        \includegraphics[trim={0.15cm, 0.15cm, 0.15cm, 0.95cm},clip,width=0.99\linewidth]{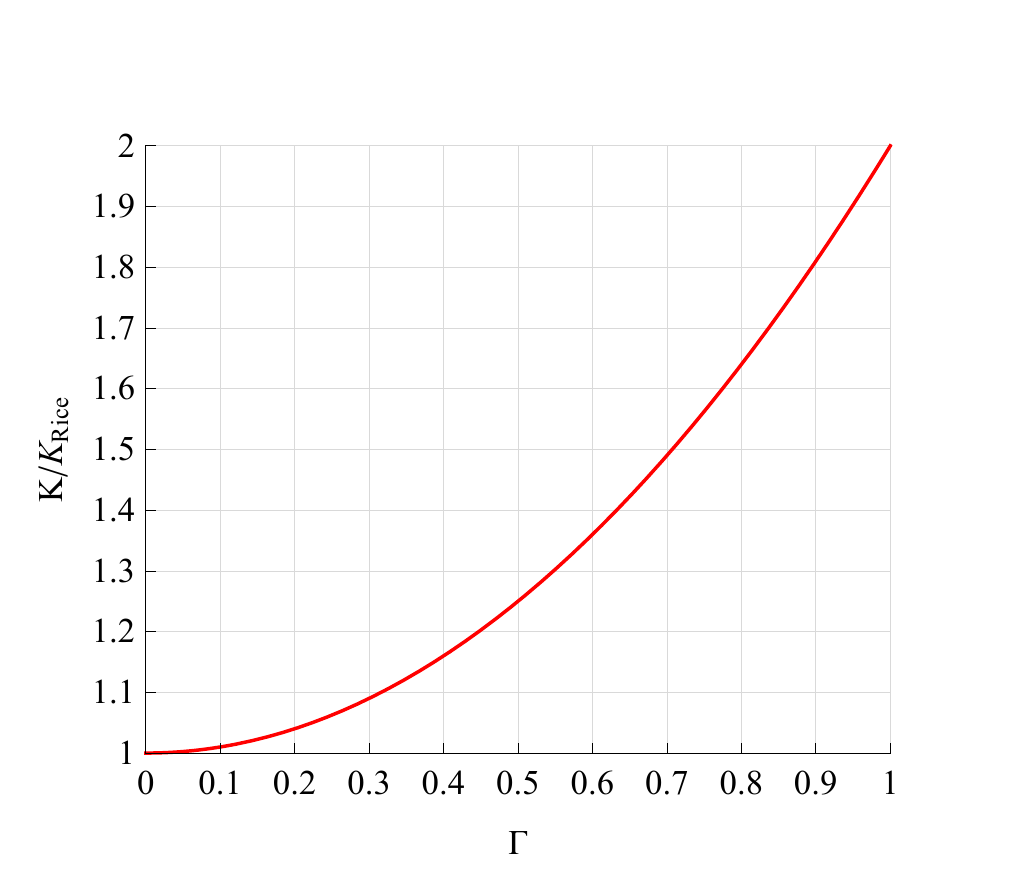}
        \caption{Dependence of $K/K_{Rice}$ on $\Gamma$}
        \label{Fig.2b}
        \vspace{4ex}
    \end{minipage}
\end{figure}

Fig.~\ref{Figure01} illustrates the functional dependence of parameter $\Delta$ versus $V_2/V_1$ (\ref{eq_2007}). In the same figure, the linear dependence of $\Gamma$ on $V_2/V_1$ is also illustrated as a benchmark.
From Fig.~\ref{Figure01} it is evident that for \mbox{$0<V_2/V_1<1$}, $\Delta$ differs $\Gamma$ not only in value, but also in terms of the character of their functional dependence on $V_2/V_1$. Consequently, when $V_2/V_1$ changes from $0.6$ to $1$, $\Delta$ changes only between $0.9$ and $1$. In general, for $0<V_2/V_1<1$, $\Delta$ is always greater than $\Gamma$.

Fig.~\ref{Fig.2a} and Fig.~\ref{Fig.2b} illustrate the dependencies of the normalized parameter $K$ ($K/K_{Rice}$) on $\Delta$ (\ref{eq_2009}) and $\Gamma$ (\ref{eq_2011}), which are clearly very different. 
For $\Delta \leq 0.8$, $K$ vs. $\Delta$ has a relatively small slope, while for $\Delta > 0.8$, the slope is very sharp. In contrast, for $0 \leq \Gamma \leq 1$, the change in parameter $K$ is relatively uniform. In other words, parameter $\Gamma$ does not change the character of the definition expression of parameter $K$ (see~(\ref{eq_2009})), while parameter $\Delta$ completely changes its character (see~(\ref{eq_2011})).  

As a consequence, parameterization based on a nonlinear relationship between $V_1$ and $V_2$ causes anomalies in graphical representations of PDF and ASEP expressions. Namely, corresponding ASEP curves are indistinguishably dense spaced for $\Delta<0.5$, which can be clearly observed from~\cite[Fig. 3]{Kim17} and~\cite[Fig. 7]{Rao14}. In addition, the shapes of the corresponding PDF curves for $\Delta<0.5$ are almost the same as the shape of a Rician PDF curve obtained for $\Delta = 0$ and the same value of $K$, which is evident from~\cite[Fig. 7]{Dur02} and~\cite[Fig. 3]{Rao14}. 
Therefore, in using $\Delta$-based parameterization, it is not possible to clearly observe the effect of the increment of $\Delta$ on PDF shape and ASEP values. Consequently, in most TWDP literature, PDF and ASEP curves are plotted only for specific values of $\Delta$, i.e. $\Delta = 0.5$ and $\Delta = 1$, for which the mentioned differences can be easily distinguished, thus avoiding graphical presentation and explanation of the results for $0\leq\Delta\leq0.5$. 

Accordingly, considering conducted elaboration, TWDP fading in this paper will be characterized by parameters $K$ and $\Gamma$. 

\subsection{Envelope PDF and CDF expressions}
To provide a mathematically convenient tool for TWDP performance evaluation, alternative exact envelope PDF and CDF expressions are proposed. Namely, it is noticed that assumptions about statistical characteristics of a complex envelope in a TWDP fading channel given in~(\ref{eq_1010}) are the same as those from~\cite{Kos78, Kos96} where the sum of signal, cochannel interference, and AWGN is modeled.
However, unlike the existing approximate TWDP PDF and CDF expressions, PDF and CDF in~\cite{Kos78, Kos96} are given in the exact form. Accordingly, using~\cite[eq. (6)]{Kos78} and~\cite[eq. (12)]{Kos96} and considering adopted parameterization, we propose the following TWDP envelope PDF and CDF expressions:
\begin{equation}
	\label{eq02'}
	\begin{split}		f_R(r)&=\frac{r}{\sigma^2}\exp{\left(-\frac{r^2 }{2\sigma^2}-K\right)} \sum_{m=0}^{\infty} \varepsilon_m (-1)^m \\
        &\times I_m\left(2r\sqrt{\frac{K}{2\sigma^2}\frac{1}{1+\Gamma^2}}\right) \\ & \times I_m\left(2r \sqrt{\frac{K}{2\sigma^2}\frac{\Gamma^2}{1+\Gamma^2}}\right)
        I_m\left(2K\frac{\Gamma}{1+\Gamma^2}\right)
	\end{split}
\end{equation}

%\begin{figure}[H]
%    \centering
%    \begin{minipage}[b,trim={0.15cm, 0.15cm, 0.15cm, 0.95cm},clip]{0.99\linewidth}
%        \centering
%        \includegraphics[trim={0.15cm, 0.15cm, 0.15cm, 0.95cm},clip,width=0.99\linewidth]{(gamma,delta)=f(v1,v2).pdf}
%        \caption{Dependence of $\Delta$ and $\Gamma$ on $V_2/V_1$}
%        \label{Figure01}
%        \vspace{4ex}
%    \end{minipage}%%
%    \vspace{0.1cm}
%    \begin{minipage}[b,trim={0.15cm, 0.15cm, 0.15cm, 0.95cm},clip]{0.99\linewidth}
%        \centering
%        \includegraphics[trim={0.15cm, 0.15cm, 0.15cm, 0.95cm},clip,width=0.99\linewidth]{K=f(delta).pdf}
%        \caption{Dependence of $K/K_{Rice}$ on $\Delta$}
%        \label{Fig.2a}
%        \vspace{4ex}
%    \end{minipage}%%
%    \vspace{0.1cm}
%    \begin{minipage}[b,trim={0.15cm, 0.15cm, 0.15cm, 0.95cm},clip]{0.99\linewidth}
%        \centering
%        \includegraphics[trim={0.15cm, 0.15cm, 0.15cm, 0.95cm},clip,width=0.99\linewidth]{K=f(gamma).pdf}
%        \caption{Dependence of $K/K_{Rice}$ on $\Gamma$}
%        \label{Fig.2b}
%        \vspace{4ex}
%    \end{minipage}
%\end{figure}

\noindent
and
\noindent
\begin{equation}
	\label{eq04}
	\begin{split}
		F_R(r) &= \frac{r^2}{2\sigma^2}\exp{\left(-\frac{r^2}{2\sigma^2}\right)} \sum_{m=0}^{\infty} \frac{(-1)^m}{m!} \left(\frac{K}{1+\Gamma^2}\right)^{m}\\
		&\times   {}_1 F_1\left(1-m;2;\frac{r^2}{2\sigma^2}\right) {}_2 F_1\left(-m,-m;1;\Gamma^2\right)  
	\end{split}
\end{equation}
\noindent where $0 \leq V_2 \leq V_1$, $\varepsilon_0 = 1$, $\varepsilon_m = 2$ for $m \geq 1$, $I_{\nu}(\cdot)$ is a modified $\nu$-th order Bessel function of the first kind, while ${}_1 F_1(\cdot;\cdot;\cdot)$ and ${}_2 F_1(\cdot,\cdot;\cdot;\cdot)$ are confluent and Gaussian hypergeometric functions, respectively.

\subsubsection{Special cases of a TWDP model}
It is easy to show that (\ref{eq02'}) and (\ref{eq04}) can be reduced to Rayleigh and Rician PDF and CDF expressions. 

The Rayleigh model assumes the absence of specular and the presence of only diffuse multipath components. It can be obtained from TWDP fading for $V_1 = V_2 = 0$, i.e. $K = 0$. So, by applying $K = 0$ into (\ref{eq02'}) and (\ref{eq04}), with $I_{\nu}(0) = 0$ for $\nu \neq 0$ and $I_{0}(0) = 1$, (\ref{eq02'}) and (\ref{eq04}) can be reduced to Rayleigh PDF and CDF expressions:
\begin{equation}
    \begin{split}
    f_R(r)\bigg\vert_{K=0}&=\frac{r}{\sigma^2}\exp{\left(-\frac{r^2 }{2\sigma^2}\right)}
    %F_R(r)\bigg\vert_{K=0}&= \frac{r^2}{2\sigma^2}\exp{\left(-\frac{r^2}{2\sigma^2}\right)}
    \end{split}
\end{equation}
\begin{equation}
    \begin{split}
    %f_R(r)\bigg\vert_{K=0}&=\frac{r}{\sigma^2}\exp{\left(-\frac{r^2 }{2\sigma^2}\right)}\\
    F_R(r)\bigg\vert_{K=0}&= \frac{r^2}{2\sigma^2}\exp{\left(-\frac{r^2}{2\sigma^2}\right)}
    \end{split}
\end{equation}
respectively. 

Rician fading assumes the presence of one specular component and many diffuse components. It can be obtained from TWDP fading for $V_2 = 0$, i.e. $\Gamma = 0$. In this case, (\ref{eq02'}) can be reduced to a well-known Rician PDF expression: \begin{equation}
f_R(r)\bigg\vert_{\Gamma=0} =\frac{r}{\sigma^2}\exp{\left(-\frac{r^2 }{2\sigma^2}-K\right)}  I_0\left(r\sqrt{2K\sigma^2}\right) 
\end{equation}

\noindent Additionally, by inserting $\Gamma = 0$ into (\ref{eq04}) and considering that
$_2F_1(\cdot,\cdot;\cdot;0) = 1$ and ${}_1 F_1(1;2;x) = (e^x-1)/x$, TWDP CDF reduces to: 
\begin{equation}
    \begin{split}
    \label{eq1022}       F_R(r)\bigg\vert_{\Gamma=0} &= 1 - \exp{\left(-\frac{r^2}{2 \sigma^2}\right)} + \frac{r^2}{2\sigma^2} \exp{\left(-\frac{r^2}{2\sigma^2}\right)} \\ & \times \sum_{m=1}^{\infty} \frac{(-1)^m}{m!} K^m {}_1 F_1\left(1-m;2;\frac{r^2}{2\sigma^2}\right) 
    \end{split}
\end{equation}
which, according to~\cite[eq. (8.352.1)]{Gra07},~\cite[eq. (8.972.1)]{Gra07} and~\cite[eq. (12)]{And10}, takes the well-known form of a Rician CDF, expressed in terms of the first-order Marcum Q-function $Q_1(\cdot)$~\cite{Rao14}:
\begin{equation}
    F_R(r)\bigg\vert_{\Gamma=0} = 1 -  Q_1\left(\sqrt{2K}, \frac{r}{\sigma}\right),
\end{equation}

\subsubsection{Convergence analysis}
It is also easy to show that (\ref{eq02'}) and (\ref{eq04}), as infinite-series expressions, are convergent.

To prove convergence of (\ref{eq02'}), d'Alembert's ratio test is used. According to the test, the infinite-series $\sum_{k} c_k$ is convergent if the limiting expression $\lim_{k\to \infty}|c_{k+1}/{c_k}|$ is smaller than~$1$. Thus, the ratio test applied to (\ref{eq02'}) yields the following expression:
\begin{equation}
\label{eq_1001}
	\begin{split}
   &\lim_{k\to \infty} \bigg| \frac{c_{k+1}}{c_k}\bigg|
   =\lim_{k\to \infty} \Bigg[
   \frac{I_{k+1}\left(2r\sqrt{\frac{K}{2\sigma^2}\frac{1}{1+\Gamma^2}}\right)}{I_{k}\left(2r\sqrt{\frac{K}{2\sigma^2}\frac{1}{1+\Gamma^2}}\right)}\\
   & \times \frac{I_{k+1}\left(2r \sqrt{\frac{K}{2\sigma^2}\frac{\Gamma^2}{1+\Gamma^2}}\right) I_{k+1}\left(2K\frac{\Gamma}{1+\Gamma^2}\right)}{I_{k}\left(2r \sqrt{\frac{K}{2\sigma^2}\frac{\Gamma^2}{1+\Gamma^2}}\right) I_{k}\left(2K\frac{\Gamma}{1+\Gamma^2}\right)}\Bigg]
   \end{split}
\end{equation}
which can be calculated using~\cite[eq. (3.12)]{Jos91} as:
\begin{equation}
\label{eq_1003}
	\begin{split}
   &\lim_{k\to \infty} \bigg| \frac{c_{k+1}}{c_k}\bigg|
   =\lim_{k\to \infty} \Bigg[
   \frac{\left(2r\sqrt{\frac{K}{2\sigma^2}\frac{1}{1+\Gamma^2}}\right)}{\left(2r\sqrt{\frac{K}{2\sigma^2}\frac{1}{1+\Gamma^2}} + k\right)}\\
   & \times \frac{\left(2r\sqrt{\frac{K}{2\sigma^2}\frac{\Gamma^2}{1+\Gamma^2}}\right)}{\left(2r\sqrt{\frac{K}{2\sigma^2}\frac{\Gamma^2}{1+\Gamma^2}} + k\right)}
    \frac{\left(2K\frac{\Gamma}{1+\Gamma^2}\right)}{\left(2K\frac{\Gamma}{1+\Gamma^2}+k\right)}
   \Bigg] = 0 < 1
	\end{split}
\end{equation}
The above expression shows that the series in (\ref{eq02'}) is convergent.

Similarly, the convergence of CDF (\ref{eq04}) is also proven using d'Alambert's ratio test, with its $k^{th}$ term  denoted by $c_k$. Since ${}_2 F_1\left(-k,-k;1;\Gamma^2\right)$ is $k^{th}$ order polynomial, due to~\cite[eq. (8.822-4), (8.911-1), (8.917.1)]{Gra07}, it can be written as $((2k)!(1+\Gamma^2)^k)/(2^k(k!)^2) + O(x^{k-1})$. Furthermore, following~\cite[eq. (8.970-1)(8.972-1)]{Gra07}, it is evident that ${}_1 F_1\left(1-k;2;\frac{r^2}{2\sigma^2}\right)$ is also a $k^{th}$ order polynomial dominated by $1/k$ when $r^2\leq2\sigma^2$ and by $[((-r^2)/(2\sigma^2))^{(k-1)}]/k!$ when $r^2>2\sigma^2$. Considering the above, d’Alambert’s ratio test yields:
\[
    \lim_{k\to \infty} \bigg| \frac{c_{k+1}}{c_k}\bigg|= 
\begin{cases}
    \lim_{k\to \infty}\left(K\frac{r^2}{2\sigma^2}\frac{(2k+1)}{(k+1)^3}\right),& r^2>2\sigma^2\\
    \lim_{k\to \infty}\left(K\frac{(2k+1)k}{(k+1)^3}\right),& r^2\leq2\sigma^2
\end{cases}
\]
which is always equal to zero and thus smaller than one. Therefore, the series in (\ref{eq04}) is also convergent.   
\subsubsection{Graphical results}
In order to investigate the accuracy of (\ref{eq02'}) and (\ref{eq04}) and their applicability for modeling various fading conditions, equations (\ref{eq02'}) and (\ref{eq04}) are plotted  for different sets of TWDP parameters. 
%After demonstrating convergence properties of (\ref{eq02'}) and (\ref{eq04}) and their possibility of reduction to Rayleigh and Rician distributions, proposed expressions are illustrated for various combinations of TWDP parameters. 

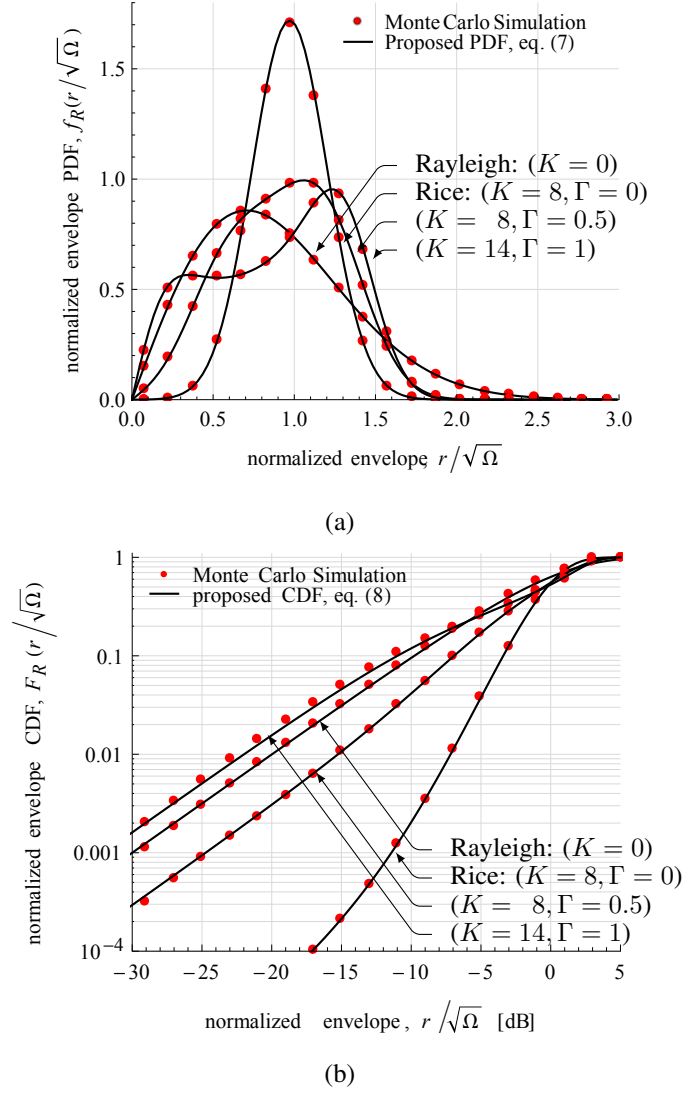
\begin{figure*}[t]
  \centering
    \begin{subfigure}{0.53\linewidth}
    %\centering
      %\includegraphics[trim={0.0cm, 0.2cm, 0.1cm, 0.3cm},clip,width=0.95\linewidth]{PDF.pdf}
      \begin{tikzpicture}[->]
            \coordinate (a) at (1.46,1.075);
            \coordinate (b) at ($ (a) + (1.8*2.3-0.2*2.3,1.0*2.3*1.3594) $);
            \node[anchor=south west,inner sep=0] at (0,0) {\includegraphics[trim={0.0cm, 0.15cm, 0.1cm, 0.3cm},clip,width=0.95\linewidth]{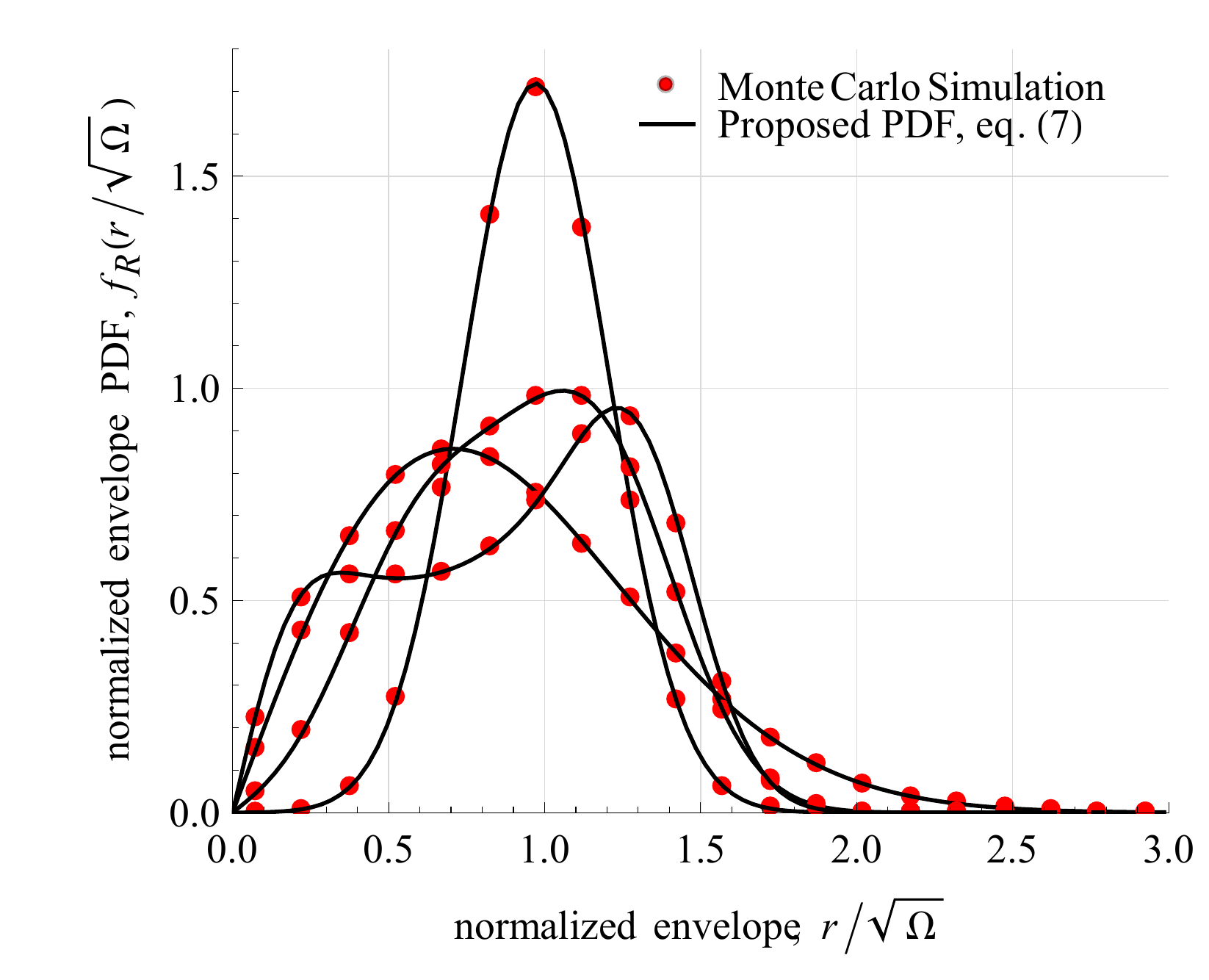}};
            \draw[black,thin,rounded corners = 2pt] ($ (b) + (0*2.3,0*2.3*1.3594) $) node[right,near start abs] {\small Rayleigh: ($K=0$)} -- ($ (b) - (0.1*2.3,0*2.3*1.3594) $)  -- ($ (b) - (0.1*2.3,0*2.3*1.3594) - (0.37*2.3,0.37*2.3*1.3594)$);
            
            \coordinate (b) at ($ (b) + (0*2.3,-0.12*2.3*1.3594) $);
            \draw[black,thin,rounded corners = 2pt] ($ (b) + (0*2.3,0*2.3*1.3594) $) node[right,near start abs] {\small Rice: ($K=8, \Gamma=0$)} -- ($ (b) - (0.1*2.3,0*2.3*1.3594) $) -- ($ (b) - (0.1*2.3,0*2.3*1.3594) - (0.21*2.3,0.21*2.3*1.3594)$);
            
            \coordinate (b) at ($ (b) + (0*2.3,-0.12*2.3*1.3594) $);
            \draw[black,thin,rounded corners = 2pt] ($ (b) + (0*2.3,0*2.3*1.3594) $) node[right,near start abs] {\small ($K=\phantom{1}8, \Gamma=0.5$)} -- ($ (b) - (0.1*2.3,0*2.3*1.3594) $) -- ($ (b) - (0.1*2.3,0*2.3*1.3594) - (0.13*2.3,0.13*2.3*1.3594)$);
            
            \coordinate (b) at ($ (b) + (0*2.3,-0.12*2.3*1.3594) $);
            \draw[black,thin,rounded corners = 2pt] ($ (b) + (0*2.3,0*2.3*1.3594) $) node[right,near start abs] {\small ($K=14, \Gamma=1$)} -- ($ (b) - (0.1*2.3,0*2.3*1.3594) $) -- ($ (b) - (0.1*2.3,0*2.3*1.3594) - (0.04*2.3,0.04*2.3*1.3594)$);
      \end{tikzpicture}
      \caption{ }
    \label{FigurePDF}
    \end{subfigure}\\
    \begin{subfigure}{0.53\linewidth}
        \centering
        \begin{tikzpicture}[->]
            \coordinate (a) at (1.50,1.04);
            \coordinate (b) at ($ (a) + (23*0.19846-0.2*2.3,1*1.40) $);
            \node[anchor=south west,inner sep=0] at (0,0) {\includegraphics[trim={0.0cm, 0.1cm, 0.1cm, 0.3cm},clip,width=0.95\linewidth]{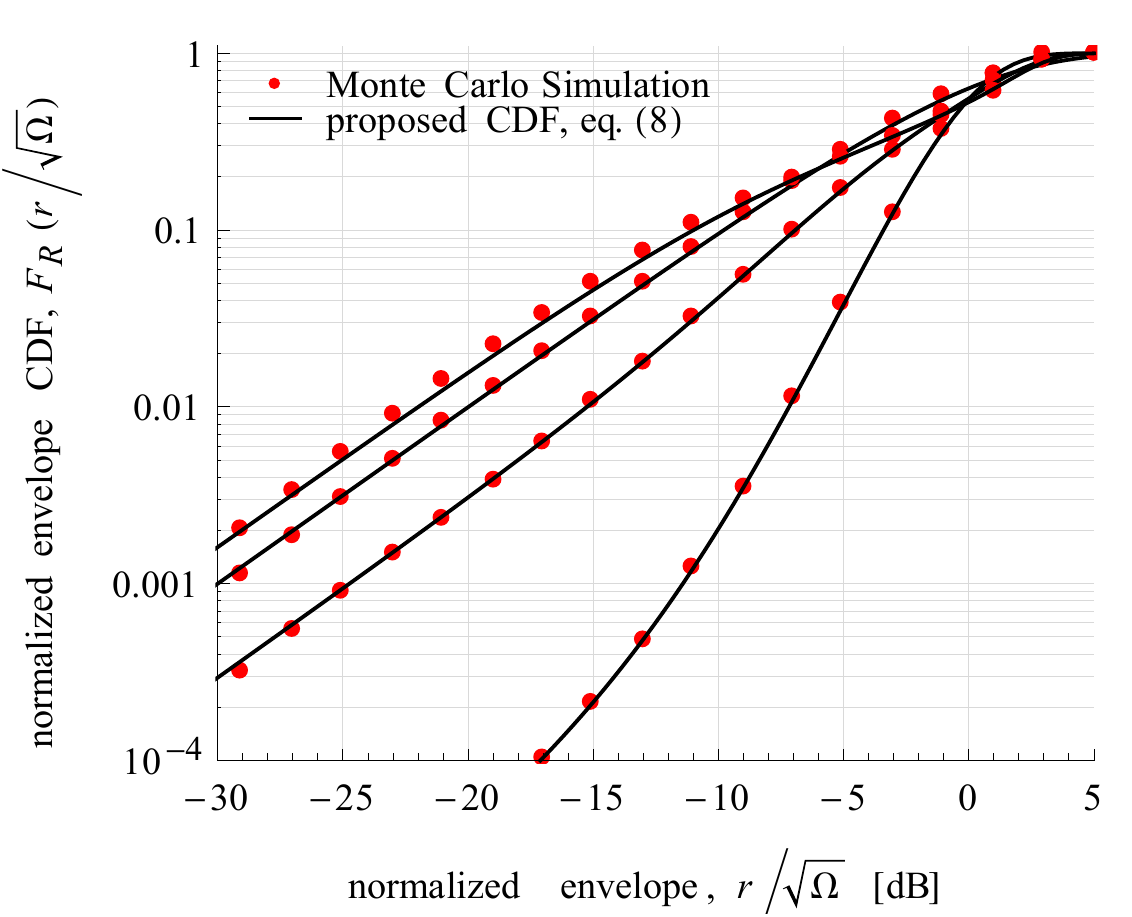}};
            \draw[black,thin,rounded corners = 2pt] ($ (b)$) node[right,near start abs] {\small Rayleigh: ($K=0$)} -- ($ (b) - (0.1*2.3,0*2.3*1.3594) $) -- ($ (b) - (0.1*2.3,0*2.3*1.3594) - (0.55*2.3,-0.55*2.3*1.3594)$);
            
            \coordinate (b) at ($ (b) + (0*2.3,-0.12*2.3*1.3594) $);
            \draw[black,thin,rounded corners = 2pt] ($ (b) + (0*2.3,0*2.3*1.3594) $) node[right,near start abs] {\small Rice: ($K=8, \Gamma=0$)} -- ($ (b) - (0.1*2.3,0*2.3*1.3594) $) -- ($ (b) - (0.1*2.3,0*2.3*1.3594) - (0.11*2.3,-0.11*2.3*1.3594)$);
            
            \coordinate (b) at ($ (b) + (0*2.3,-0.12*2.3*1.3594) $);
            \draw[black,thin,rounded corners = 2pt] ($ (b) + (0*2.3,0*2.3*1.3594) $) node[right,near start abs] {\small ($K=\phantom{1}8, \Gamma=0.5$)} -- ($ (b) - (0.1*2.3,0*2.3*1.3594) $) -- ($ (b) - (0.1*2.3,0*2.3*1.3594) - (0.57*2.3,-0.57*2.3*1.3594)$);
            
            \coordinate (b) at ($ (b) + (0*2.3,-0.12*2.3*1.3594) $);
            \draw[black,thin,rounded corners = 2pt] ($ (b) + (0*2.3,0*2.3*1.3594) $) node[right,near start abs] {\small ($K=14, \Gamma=1$)} -- ($ (b) - (0.1*2.3,0*2.3*1.3594) $) -- ($ (b) - (0.1*2.3,0*2.3*1.3594) - (0.85*2.3,-0.85*2.3*1.3594)$);
      \end{tikzpicture}
        \caption{ }
    \label{FigureCDF}
    \end{subfigure}  
  \caption{TWDP normalized envelope (a) PDF and (b) CDF curves for various combinations of~$K$~and~$\Gamma$}
  \label{Fig.3}
\end{figure*}

Equation~(\ref{eq02'}) is used to plot the normalized envelope PDF, $f_R(r/\sqrt{\Omega})$, for different fading conditions: 
Rician with $K = 8$ and $\Gamma = 0$; Rayleigh with $K = 0$; and others, with $K = 8$ and $\Gamma = 0.5$; and $K = 14$ and $\Gamma = 1$.
%, where $\Omega$ represents the average signal power, i.e., $\Omega=V_1^2+V_2^2+2\sigma^2$.
Fig.~\ref{FigurePDF} depicts these curves together with corresponding normalized histograms created by Monte Carlo simulation. All curves are obtained by limiting truncation error below $10^{-6}$, i.e., by employing up to $35$ summation terms in all tested cases. Each normalized histogram, composed of $20$ equally spaced bins, is computed independently by generating $10^6$ samples for the considered fading conditions. Fig.~\ref{FigurePDF} shows matching results between the analytical and simulated approaches, thus validating the proposed PDF expression in diverse fading conditions. 

Fig. \ref{FigureCDF} compares normalized envelope CDF curves $F_R(r/\sqrt{\Omega})$ obtained from (\ref{eq04}) with normalized cumulative histograms. Similarly, Monte Carlo simulation is used to generate histograms with the same set of parameters as in the PDF comparison. Analytically obtained curves are generated by employing up to $118$ summation terms in order to achieve a truncation error of less than $10^{-26}$.
Normalized cumulative histograms are created from $10^6$ samples divided into $20$ bins.
The conducted comparison shows matching results between the analytical and simulated approaches, thus demonstrating the applicability of (\ref{eq04}) for accurate calculation of CDF values in different fading conditions. 

%{It is also interesting to note that proposed expressions reduce the time necessary for calculation of CDF values in respect to Monte Carlo simulation. Accordingly, (\ref{eq02'}) and (\ref{eq04})  provide both accurate and computationally efficient tool for TWDP fading channel performance evaluation.}

\section{Alternative form of TWDP SNR MGF expression}
\label{sec:III}
In this section, the alternative form of the MGF of the SNR is derived based on the proposed CDF expression. 
Here, the well-known relationship between CDF and MGF is used~\cite[eq. (1.2)]{Alo}: \begin{equation}
	\label{eq06}
	\begin{split}
		\mathcal{M}_\gamma (s)&=\int_0^{+\infty}f_\gamma (\gamma)  \exp{\left(s\gamma\right)} \diff{\gamma}\\
        &=\mathcal{L}\left\{ f_\gamma(\gamma);\gamma,-s\right\}\\
        %&=\int_0^{+\infty} \frac{\diff{}}{\diff{\gamma}} F_\gamma (\gamma)  \exp{\left(s\gamma\right)} \diff{\gamma}\\
        &=\mathcal{L}\left\{ \frac{\diff{}}{\diff{\gamma}} F_\gamma(\gamma);\gamma,-s\right\}\\
        &=-s \mathcal{L}\left\{ F_\gamma(\gamma);\gamma,-s\right\} - F_\gamma (\gamma=0)\\
        &=-s \mathcal{L}\left\{ F_\gamma(\gamma);\gamma,-s \right\}
	\end{split}
\end{equation}
where $\mathcal{L}\{h(t);t,p\}\triangleq\int_{0}^{\infty}h(t) e^{-p t}\diff{t}$ represents Laplace transform of $h(t)$ from $t$-domain into the $p$-domain, and $F_\gamma(\gamma)$ is the CDF of the SNR. $F_\gamma(\gamma)$ is obtained from (\ref{eq04}) according to the random variable transformation $\gamma=r^2\frac{E_s}{N_0}$, as:
\begin{equation}
	\label{eq05}
	\begin{split}
		F_\gamma(\gamma)&=\frac{\gamma}{\gamma_0}\left(1+K\right)		 \exp\left({-\frac{\gamma}{\gamma_0}(1+K)}\right) \sum_{m=0}^{\infty}
        \frac{(-1)^m}{m!} \\
        &\times \left(\frac{K}{1 + \Gamma^2}\right)^{m} {}_1 F_1\left(1-m;2;\frac{\gamma}{\gamma_0}(1+K)\right) \\
         &\times {}_2 F_1\left(-m,-m;1;\Gamma^2\right)
	\end{split}
\end{equation}
where \mbox{$\gamma_0=2\sigma^2 (1+K)\frac{E_s}{N_0}$} is the average SNR, $E_s$ denotes symbol energy, and $N_0/2$ is the power spectral density of the white Gaussian noise.

For simplicity, (\ref{eq05}) is expressed in the following form:
\begin{equation}
	\label{eq_06}
	\begin{split}
		F_\gamma(\gamma)=\sum_{m=0}^{\infty} 
        A \gamma B_m \exp\left({-A \gamma}\right) {}_1 F_1\left(1-m;2;A \gamma\right)
	\end{split}
\end{equation}
where $B_m= \left(-K/(1+\Gamma^2)\right)^m
        {}_2 F_1\left(-m,-m;1;\Gamma^2\right)/m!$ and \mbox{$A=(1+K)/\gamma_0$}.
Based on~\cite[eq. (07.20.16.0001.01)]{Wol}, (\ref{eq_06}) is further simplified as:
 \begin{equation}
     \label{eq_10}
     F_\gamma(\gamma)=\sum_{m=0}^{\infty} A \gamma B_m  {}_1 F_1\left(1+m;2;-A \gamma\right)
 \end{equation}
Laplace transform of~(\ref{eq_10}) is then obtained using \cite[eq. (3.35.1-2)]{Prud_Lap} as:
\begin{equation}
    \mathcal{L}\left\{ F_\gamma(\gamma);\gamma,s\right\} = \sum_{m=0}^{\infty} 
        \frac{A B_m}{s^2} {}_2 F_1\left(1+m;2;2,-\frac{A}{s} \right)
\end{equation}
which, according to~\cite[eq. (07.23.03.0080.01)]{Wol}, can be expressed in the following form:
   \begin{equation}
	\label{eq_07}
	\begin{split}
		\mathcal{L}\left\{ F_\gamma(\gamma);\gamma,s\right\}=\sum_{m=0}^{\infty} 
         \frac{A B_m}{\left(A+s\right)^2}\left(\frac{s}{A+s}\right)^{m-1}
	\end{split}
\end{equation}
Finally, by combining (\ref{eq06}) and (\ref{eq_07}), the MGF is derived as:
\begin{equation}
	\label{eq_08}
	\begin{split}
		\mathcal{M}_\gamma(s) &= 
		\frac{1+K}{1+K-s\gamma_0}
		\sum_{m=0}^{\infty} \frac{1}{m!} \left(\frac{K}{1+\Gamma^2}\right)^m \\ &\times \left(\frac{\gamma_0 s}{1+K-s\gamma_0}\right)^{m}   {}_2F_1\left(-m,-m;1;\Gamma^2\right)
	\end{split}
\end{equation}
which represents an alternative form of the exact TWDP MGF of the SNR. 

It can be proven that (\ref{eq_08}) can be easily transformed into the well-known TWDP MGF expression form~\cite[eq. (25)]{Rao15} (originally given in terms of $K$ and $\Delta$). Namely, by using the identity between the Gaussian hypergeometric function and the Legendre polynomial given by~\cite[eq. (15.4.14)]{Abr72}, as well as the identity between the Legendre polynomial and the first-kind zero-order Bessel function given by~\cite[eq. (0.6)]{Koepf1998}, and after some simple manipulations, it can be shown that:
\begin{equation}
\label{eq_100}
    \sum_{m=0}^{\infty} \frac{1}{m!} a^m {}_2F_1(-m,-m;1;b) = \exp{\left(a+ab\right)} I_0\left(2a\sqrt{b}\right)
\end{equation}
Therefore, by using (\ref{eq_100}), (\ref{eq_08}) can be written as:
\begin{equation}
    \begin{split}
    \label{eq_09}
        \mathcal{M}_\gamma(s) &= \frac{1+K}{1+K-\gamma_0 s}           \exp{\left(\frac{\gamma_0 K s}{1+K-\gamma_0 s}\right)}\\ &\times I_0\left(\frac{2 \Gamma \gamma_0 K s }{(1+K-\gamma_0 s)(1+\Gamma^2)}\right)
    \end{split}
\end{equation}
which is the same expression as the verified SNR MGF from~\cite{Rao15}, only expressed in terms of $K$ and $\Gamma$.

Although simple, the analytical form of MGF expressed by (\ref{eq_09}) has not been often used for error rate performance evaluation in TWDP fading channels. The main disadvantage with this expression is its unfavorable analytical form for mathematical manipulations. 
In contrast, the analytical form of MGF as expressed by (\ref{eq_08}) enables derivation of the exact expressions for the performance evaluation in a variety of TWDP fading conditions. 

\section{Error probability of M-ary PSK receiver in TWDP fading channel}
\label{sec:IV}
\subsection{The exact M-ary PSK ASEP expression}
This section demonstrates the applicability of the proposed TWDP SNR MGF (\ref{eq_08}) for derivation of the exact M-ary PSK ASEP expression, where M represents the order of PSK modulation.

M-ary PSK ASEP in a TWDP fading channel can be determined from~\cite[eq. (5.78)]{Alo}:
\begin{equation}
\label{eq_20}
    P_s(\gamma_0)=\frac{1}{\pi}\int_{0}^{\pi-\frac{\pi}{M}} \mathcal{M}_\gamma\left(-\frac{\sin^2{\frac{\pi}{M}}}{\sin^2{\theta}}\right) \diff{\theta}
\end{equation}
where $\mathcal{M}_\gamma\left(\cdot\right)$ represents the MGF of the SNR given in (\ref{eq_08}). Accordingly, equation (\ref{eq_20}) can be expressed as \mbox{$P_s(\gamma_0) =
        2\mathcal{I} \vert_{0}^{\frac{\pi}{2}}-\mathcal{I} \vert_{0}^{\frac{\pi}{M}}$}, 
where $\mathcal{I}$ represents the indefinite integral defined as:   
\begin{equation}
    \begin{split}
    \label{eq_21}
        \mathcal{I}&=\frac{1}{\pi} \int \mathcal{M}_\gamma\left(-\frac{\sin^2{\frac{\pi}{M}}}{\sin^2{\theta}}\right) \diff{\theta}\\
        &=\frac{1}{\pi} \sum_{m=0}^{\infty} 
        \frac{1}{m!}
        \left(\frac{K}{1+\Gamma^2}\right)^m
         {}_2F_1\left(-m,-m;1;\Gamma^2\right) \\
        & \times \int \left(\frac{1+K}{1+K-\gamma_0 s}\right) \left(\frac{ \gamma_0 s }{1+K-\gamma_0 s}\right)^{m} \diff{\theta}\Bigg\vert_{s=-\frac{\sin^2{\frac{\pi}{M}}}{\sin^2{\theta}}}        %\sin^2{\pi/M}
    \end{split}
\end{equation}
which can be solved using Wolfram Mathematica as:
\begin{equation}
    \begin{split}
    \label{eq_22}
        \mathcal{I}&=\frac{1}{3\pi} \frac{1+K}{\gamma_0} \sum_{m=0}^{\infty} \frac{(-1)^m}{m!}
        \left(\frac{K}{1+\Gamma^2}\right)^m \frac{\sin^3{\theta}}{\sin^2{\pi/M}}\\
        & \times {AF}_1\left(\frac{3}{2}; \frac{1}{2},1+m;\frac{5}{2};\sin^2{\theta},-\frac{1+K}{\gamma_0}\frac{\sin^2{\theta}}{\sin^2{\pi/M}}\right) \\
        & \times {}_2F_1\left(-m,-m;1;\Gamma^2\right),~~~~~\text{for } 0 \leq \theta\leq \frac{\pi}{2}
    \end{split}
\end{equation}
where $AF_1(\cdot;\cdot,\cdot; \cdot;\cdot,\cdot)$ is an Appell hypergeometric function. 
Considering the above, the integral in (\ref{eq_20}) can be solved as:
\begin{equation}
    \begin{split}
    \label{eq_23}
        P_s&(\gamma_0) =
        \frac{\sin{\frac{\pi}{M}}}{3\pi} \frac{1+K}{\gamma_0} \sum_{m=0}^{\infty}\frac{1}{m!}{}_2F_1\left(-m,-m;1;\Gamma^2\right)   \\
        &\times {\left(\frac{-K}{1+\Gamma^2}\right)^m} \left[\frac{3\pi}{2 \sin^3{\frac{\pi}{M}}} {}_2F_1\left(\frac{3}{2},1+m;2;-\frac{1+K}{\gamma_0\sin{\frac{\pi}{M}}}\right)\right.\\
        &\left.- AF_1\left(\frac{3}{2}; \frac{1}{2},1+m;\frac{5}{2};\sin^2{\frac{\pi}{M}},-\frac{1+K}{\gamma_0}\right)\right]
    \end{split}
\end{equation}
which represents M-ary PSK ASEP given as the exact analytical expression.

\subsection{Asymptotic expression of M-ary PSK ASEP}
\label{Aproksimacija}
To gain further insight into the TWDP M-ary PSK ASEP behavior, the asymptotic ASEP for large values of $\gamma_0$ is derived. Furthermore, this allow us to relax the computational complexity which occurs for large values of $K$. 

Considering that \mbox{$AF_1\left(a; b_1,b_2;c;z_1,z\right)\sim {}_2 F_1\left(a,b_1;c;z_1\right)$} and \mbox{${}_2 F_1\left(a,b;c;z\right)\sim 1$} when \mbox{$z\xrightarrow{} 0$}, equation (\ref{eq_23}) for large values of $\gamma_0$ can be expressed as:
\begin{equation}
    \begin{split}
    \label{eq_24}
        P_s(\gamma_0) &\approx \frac{\sin{\frac{\pi}{M}}}{3\pi} \frac{1+K}{\gamma_0} \sum_{m=0}^{\infty} \frac{\left(\frac{K}{1+\Gamma^2}\right)^m}{m!} {}_2F_1\left(-m,-m;1;\Gamma^2\right) \\
        & \times \left[\frac{3\pi}{2 \sin^3{\frac{\pi}{M}}} - {}_2 F_1\left(\frac{3}{2},\frac{1}{2};\frac{5}{2};\sin^2{\frac{\pi}{M}}\right)\right]
    \end{split}
\end{equation}
Equation (\ref{eq_24}) can be further simplified using the identity~\cite[p. 24]{Ste63} and equation (\ref{eq_100}), as: 
\begin{equation}
    \begin{split}
    \label{eq_23'}
        P_s(\gamma_0) \approx & \frac{1+K}{2\pi\gamma_0} \frac{\pi-\frac{\pi}{M}+\frac{1}{2}\sin{\frac{2\pi}{M}}}{\sin^2{\frac{\pi}{M}}} e^{-K} I_0\left(\frac{2\Gamma K}{1 + \Gamma^2}\right)
\end{split}
\end{equation}
which represents a simple, closed-form asymptotic M-ary PSK ASEP expression.
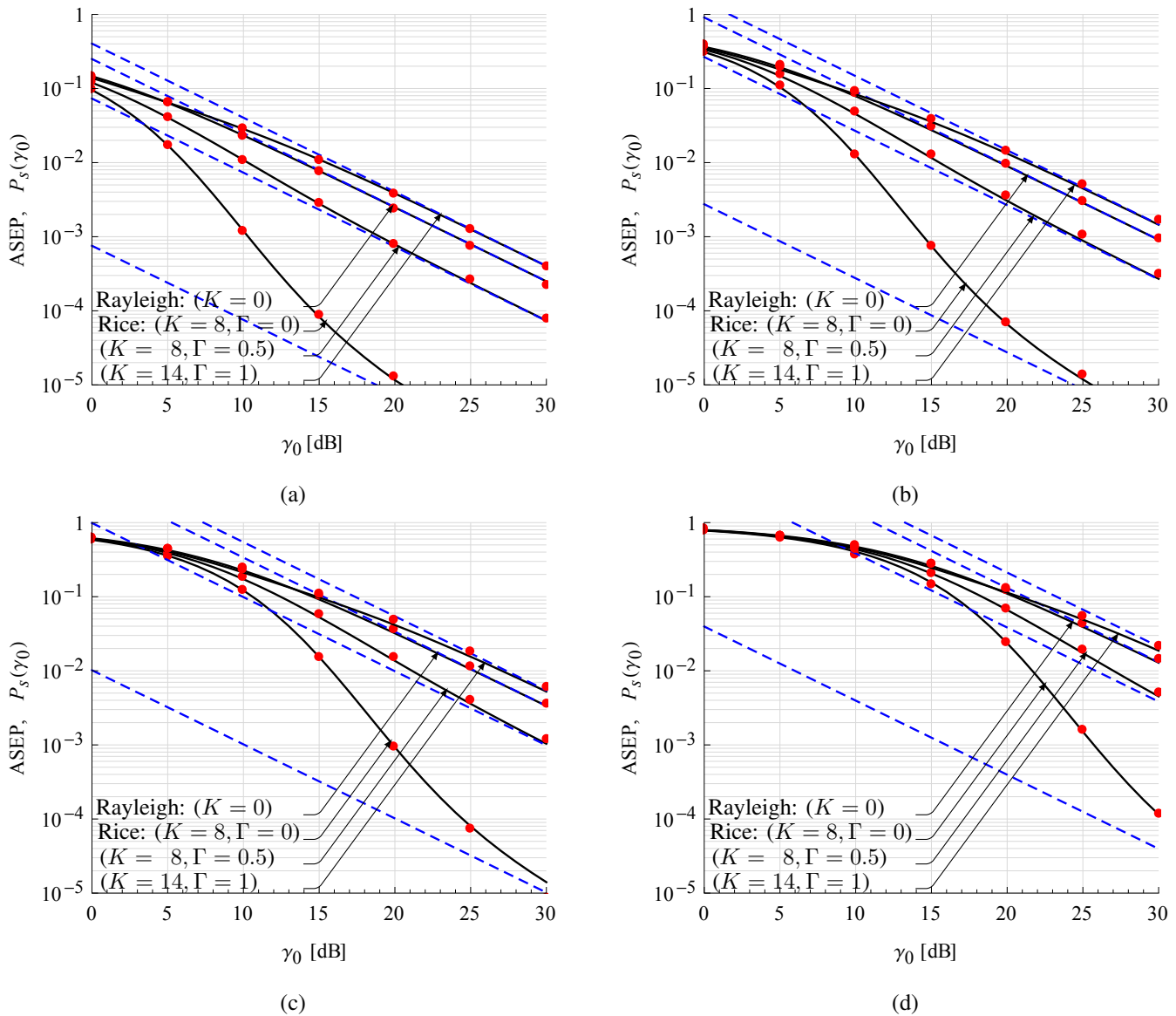
\begin{figure*}[t]
    \centering
    \hspace*{-1.0cm}
    \begin{tabular}[c]{cc}
    \begin{subfigure}[c]{0.54\linewidth}
    %\begin{subfigure}[c]{0.49\linewidth}
        %\centering
        %\includegraphics[trim={0.8cm, 0.2cm, 0.2cm, 0.3cm},clip,width=.95\linewidth]{BER2.pdf}
        \begin{tikzpicture}[->]
            \coordinate (a) at (1.4,1.075);
            %\coordinate (a) at (1.3,1.075);
            \coordinate (b) at ($ (a) + (1.2*2.3+0.2*2.3,0.4*2.3*1.3594) $);
            \node[anchor=south west,inner sep=0] at (0,0) {\includegraphics[trim={0.8cm, 0.2cm, 0.2cm, 0.3cm},clip,width=.95\linewidth]{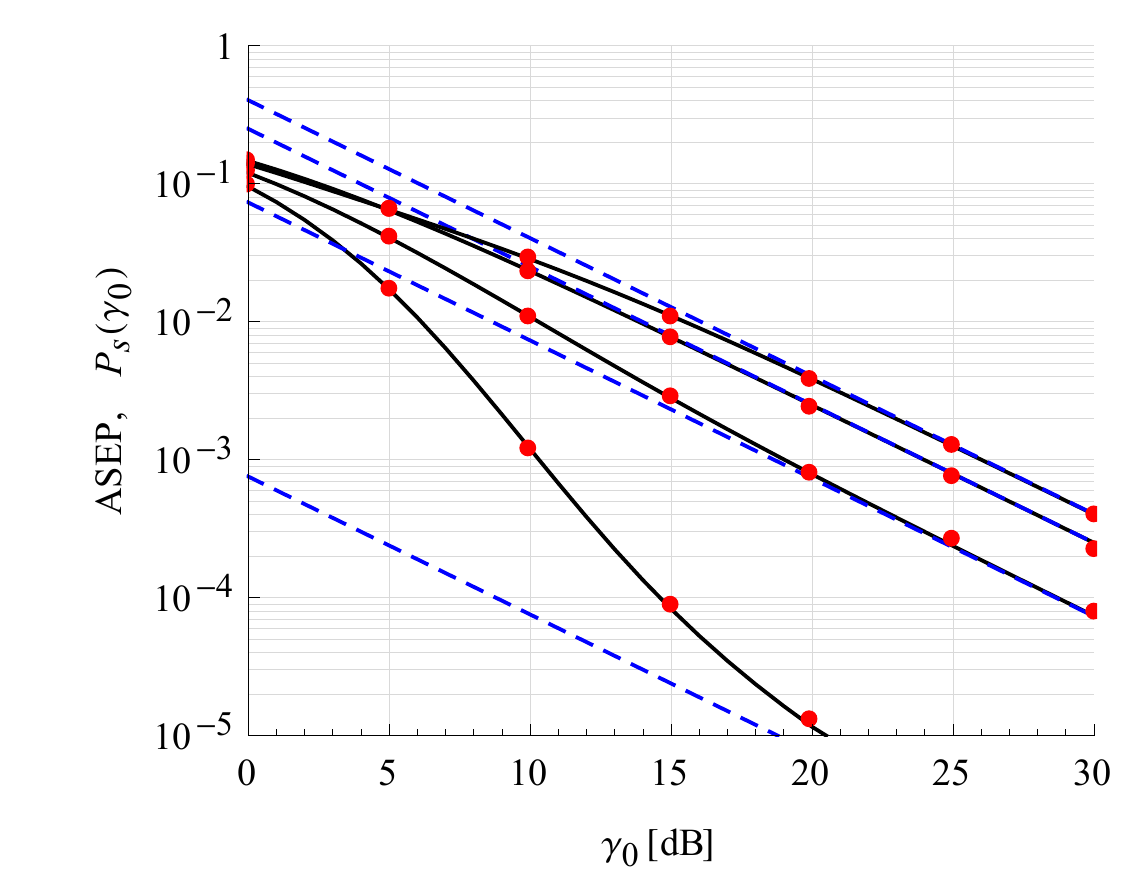}};
            \node[left,near start abs,minimum size=25mm,align = left,anchor=base east] at ($ (b) + (0*2.3,0*2.3*1.3594) $) {\small Rayleigh: ($K=0$)~~\phantom{..}};
            \draw[black,thin,rounded corners = 2pt] ($ (b) + (0*2.3,0*2.3*1.3594) $) -- ($ (b) + (0.1*2.3,0*2.3*1.3594) $) -- ($ (b) + (0.1*2.3,0*2.3*1.3594) + (0.50*2.3,0.50*2.3*1.3594)$);
            \coordinate (b) at ($ (b) + (0*2.3,-0.12*2.3*1.3594) $);
            \node[left,near start abs,minimum size=25mm,align = left,anchor=base east] at ($ (b) + (0*2.3,0*2.3*1.3594) $) {\small Rice: ($K=8, \Gamma=0$)};
            \draw[black,thin,rounded corners = 2pt] ($ (b) + (0*2.3,0*2.3*1.3594) $) -- ($ (b) + (0.1*2.3,0*2.3*1.3594) $) -- ($ (b) + (0.1*2.3,0*2.3*1.3594) + (0.06*2.3,0.06*2.3*1.3594)$);
            \coordinate (b) at ($ (b) + (0*2.3,-0.12*2.3*1.3594) $);
            \node[left,near start abs,minimum size=25mm,align = left,anchor=base east] at ($ (b) + (0*2.3,0*2.3*1.3594) $) {\small ($K=\phantom{1}8, \Gamma=0.5$)~\phantom{...}};
            \draw[black,thin,rounded corners = 2pt] ($ (b) + (0*2.3,0*2.3*1.3594) $) -- ($ (b) + (0.1*2.3,0*2.3*1.3594) $) -- ($ (b) + (0.1*2.3,0*2.3*1.3594) + (0.54*2.3,0.54*2.3*1.3594)$);
            \coordinate (b) at ($ (b) + (0*2.3,-0.12*2.3*1.3594) $);
            \node[left,near start abs,minimum size=25mm,align = left,anchor=base east] at ($ (b) + (0*2.3,0*2.3*1.3594) $) {\small ($K=14, \Gamma=1$)\phantom{.5}~\phantom{...}};
            \draw[black,thin,rounded corners = 2pt] ($ (b) + (0*2.3,0*2.3*1.3594) $) -- ($ (b) + (0.1*2.3,0*2.3*1.3594) $) -- ($ (b) + (0.1*2.3,0*2.3*1.3594) + (0.82*2.3,0.82*2.3*1.3594)$);
      \end{tikzpicture}
        \caption{ }
        %, K=\{0,8,8,14\}, $\Gamma$=\{-,0,0.5,1\}, greška=$10^{-6}$, broj članova=\{2,28,42,78\}}
        \label{ASEP2}
    \end{subfigure}&
    \begin{subfigure}[c]{0.54\linewidth}
        %\centering
        \begin{tikzpicture}[->]
            \coordinate (a) at (1.4,1.075);
            \coordinate (b) at ($ (a) + (1.2*2.3+0.2*2.3,0.4*2.3*1.3594) $);
            \node[anchor=south west,inner sep=0] at (0,0) {\includegraphics[trim={0.8cm, 0.2cm, 0.2cm, 0.3cm},clip,width=.95\linewidth]{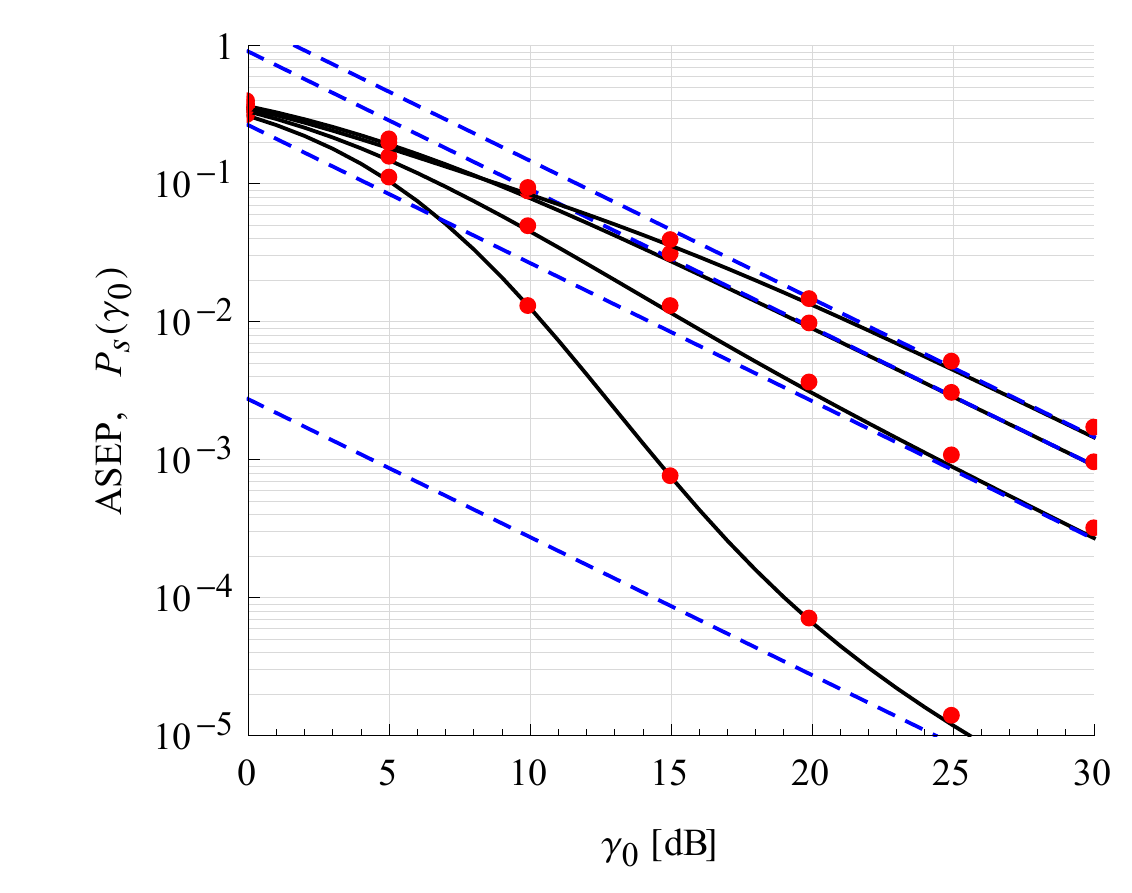}};
            \node[left,near start abs,minimum size=25mm,align = left,anchor=base east] at ($ (b) + (0*2.3,0*2.3*1.3594) $) {\small Rayleigh: ($K=0$)~~\phantom{..}};
            \draw[black,thin,rounded corners = 2pt] ($ (b) + (0*2.3,0*2.3*1.3594) $) -- ($ (b) + (0.1*2.3,0*2.3*1.3594) $) -- ($ (b) + (0.1*2.3,0*2.3*1.3594) + (0.65*2.3,0.65*2.3*1.3594)$);
            \coordinate (b) at ($ (b) + (0*2.3,-0.12*2.3*1.3594) $);
            \node[left,near start abs,minimum size=25mm,align = left,anchor=base east] at ($ (b) + (0*2.3,0*2.3*1.3594) $) {\small Rice: ($K=8, \Gamma=0$)};
            \draw[black,thin,rounded corners = 2pt] ($ (b) + (0*2.3,0*2.3*1.3594) $) -- ($ (b) + (0.1*2.3,0*2.3*1.3594) $) -- ($ (b) + (0.1*2.3,0*2.3*1.3594) + (0.24*2.3,0.24*2.3*1.3594)$);
            \coordinate (b) at ($ (b) + (0*2.3,-0.12*2.3*1.3594) $);
            \node[left,near start abs,minimum size=25mm,align = left,anchor=base east] at ($ (b) + (0*2.3,0*2.3*1.3594) $) {\small ($K=\phantom{1}8, \Gamma=0.5$)~\phantom{...}};
            \draw[black,thin,rounded corners = 2pt] ($ (b) + (0*2.3,0*2.3*1.3594) $) -- ($ (b) + (0.1*2.3,0*2.3*1.3594) $) -- ($ (b) + (0.1*2.3,0*2.3*1.3594) + (0.69*2.3,0.69*2.3*1.3594)$);
            \coordinate (b) at ($ (b) + (0*2.3,-0.12*2.3*1.3594) $);
            \node[left,near start abs,minimum size=25mm,align = left,anchor=base east] at ($ (b) + (0*2.3,0*2.3*1.3594) $) {\small ($K=14, \Gamma=1$)\phantom{.5}~\phantom{...}};
            \draw[black,thin,rounded corners = 2pt] ($ (b) + (0*2.3,0*2.3*1.3594) $) -- ($ (b) + (0.1*2.3,0*2.3*1.3594) $) -- ($ (b) + (0.1*2.3,0*2.3*1.3594) + (0.96*2.3,0.96*2.3*1.3594)$);
      \end{tikzpicture}
        \caption{ }
        %, K=\{0,8,8,14\}, $\Gamma$=\{-,0,0.5,1\}, greška=$10^{-6}$, broj članova=\{2,28,43,79\}}
        \label{ASEP4}
    \end{subfigure}\\
    \begin{subfigure}[c]{0.54\linewidth}
        %\centering
        \begin{tikzpicture}[->]
            \coordinate (a) at (1.4,1.075);
            \coordinate (b) at ($ (a) + (1.2*2.3+0.2*2.3,0.4*2.3*1.3594) $);
            \node[anchor=south west,inner sep=0] at (0,0) {\includegraphics[trim={0.8cm, 0.2cm, 0.2cm, 0.3cm},clip,width=.95\linewidth]{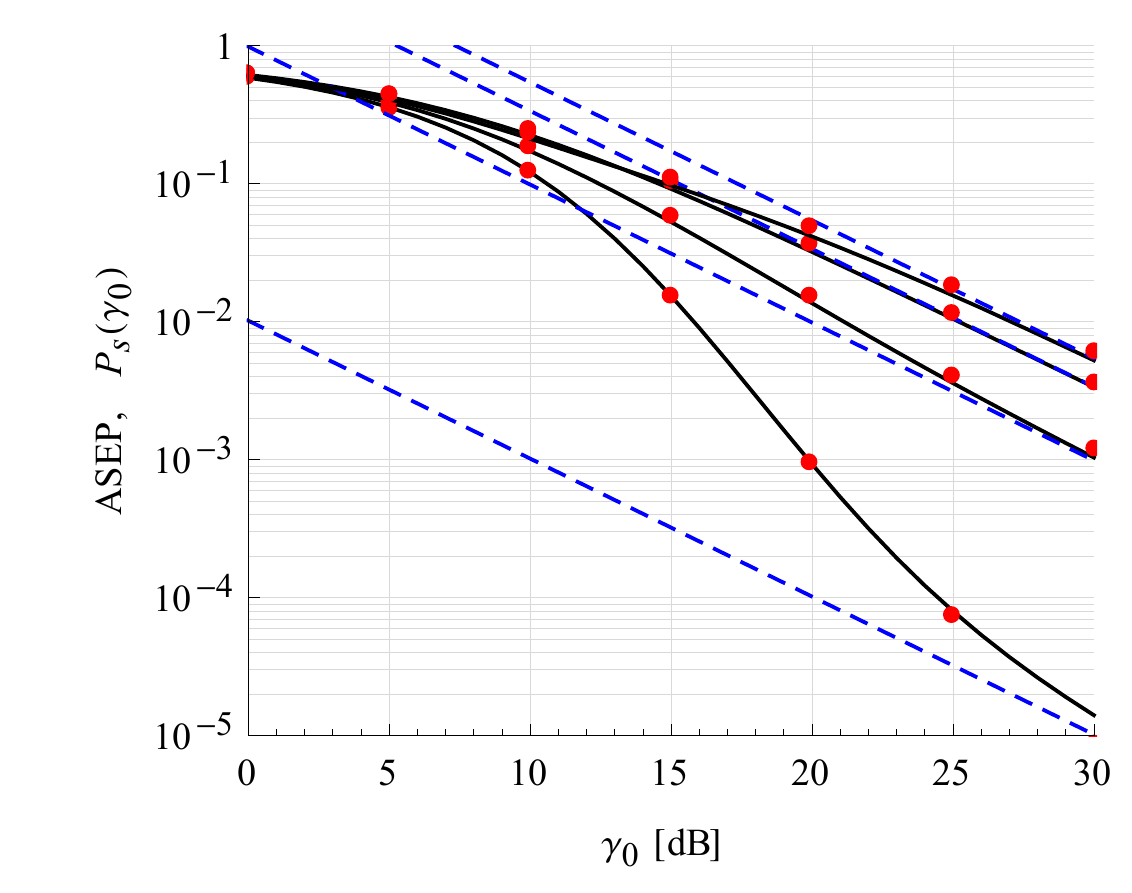}};
            \node[left,near start abs,minimum size=25mm,align = left,anchor=base east] at ($ (b) + (0*2.3,0*2.3*1.3594) $) {\small Rayleigh: ($K=0$)~~\phantom{..}};
            \draw[black,thin,rounded corners = 2pt] ($ (b) + (0*2.3,0*2.3*1.3594) $) -- ($ (b) + (0.1*2.3,0*2.3*1.3594) $) -- ($ (b) + (0.1*2.3,0*2.3*1.3594) + (0.80*2.3,0.80*2.3*1.3594)$);
            \coordinate (b) at ($ (b) + (0*2.3,-0.12*2.3*1.3594) $);
            \node[left,near start abs,minimum size=25mm,align = left,anchor=base east] at ($ (b) + (0*2.3,0*2.3*1.3594) $) {\small Rice: ($K=8, \Gamma=0$)};
            \draw[black,thin,rounded corners = 2pt] ($ (b) + (0*2.3,0*2.3*1.3594) $) -- ($ (b) + (0.1*2.3,0*2.3*1.3594) $) -- ($ (b) + (0.1*2.3,0*2.3*1.3594) + (0.49*2.3,0.49*2.3*1.3594)$);
            \coordinate (b) at ($ (b) + (0*2.3,-0.12*2.3*1.3594) $);
            \node[left,near start abs,minimum size=25mm,align = left,anchor=base east] at ($ (b) + (0*2.3,0*2.3*1.3594) $) {\small ($K=\phantom{1}8, \Gamma=0.5$)~\phantom{...}};
            \draw[black,thin,rounded corners = 2pt] ($ (b) + (0*2.3,0*2.3*1.3594) $) -- ($ (b) + (0.1*2.3,0*2.3*1.3594) $) -- ($ (b) + (0.1*2.3,0*2.3*1.3594) + (0.86*2.3,0.86*2.3*1.3594)$);
            \coordinate (b) at ($ (b) + (0*2.3,-0.12*2.3*1.3594) $);
            \node[left,near start abs,minimum size=25mm,align = left,anchor=base east] at ($ (b) + (0*2.3,0*2.3*1.3594) $) {\small ($K=14, \Gamma=1$)\phantom{.5}~\phantom{...}};
            \draw[black,thin,rounded corners = 2pt] ($ (b) + (0*2.3,0*2.3*1.3594) $) -- ($ (b) + (0.1*2.3,0*2.3*1.3594) $) -- ($ (b) + (0.1*2.3,0*2.3*1.3594) + (1.11*2.3,1.11*2.3*1.3594)$);
      \end{tikzpicture}
        \caption{ }
        %, K=\{0,8,8,14\}, $\Gamma$=\{-,0,0.5,1\}, greška=$10^{-6}$, broj članova=\{2,28,43,78\}}
        \label{ASEP8}
    \end{subfigure}
    &\begin{subfigure}[c]{0.54\linewidth}
        %\centering
        \begin{tikzpicture}[->]
            \coordinate (a) at (1.4,1.075);
            \coordinate (b) at ($ (a) + (1.2*2.3+0.2*2.3,0.4*2.3*1.3594) $);
            \node[anchor=south west,inner sep=0] at (0,0) {\includegraphics[trim={0.8cm, 0.2cm, 0.2cm, 0.3cm},clip,width=.95\linewidth]{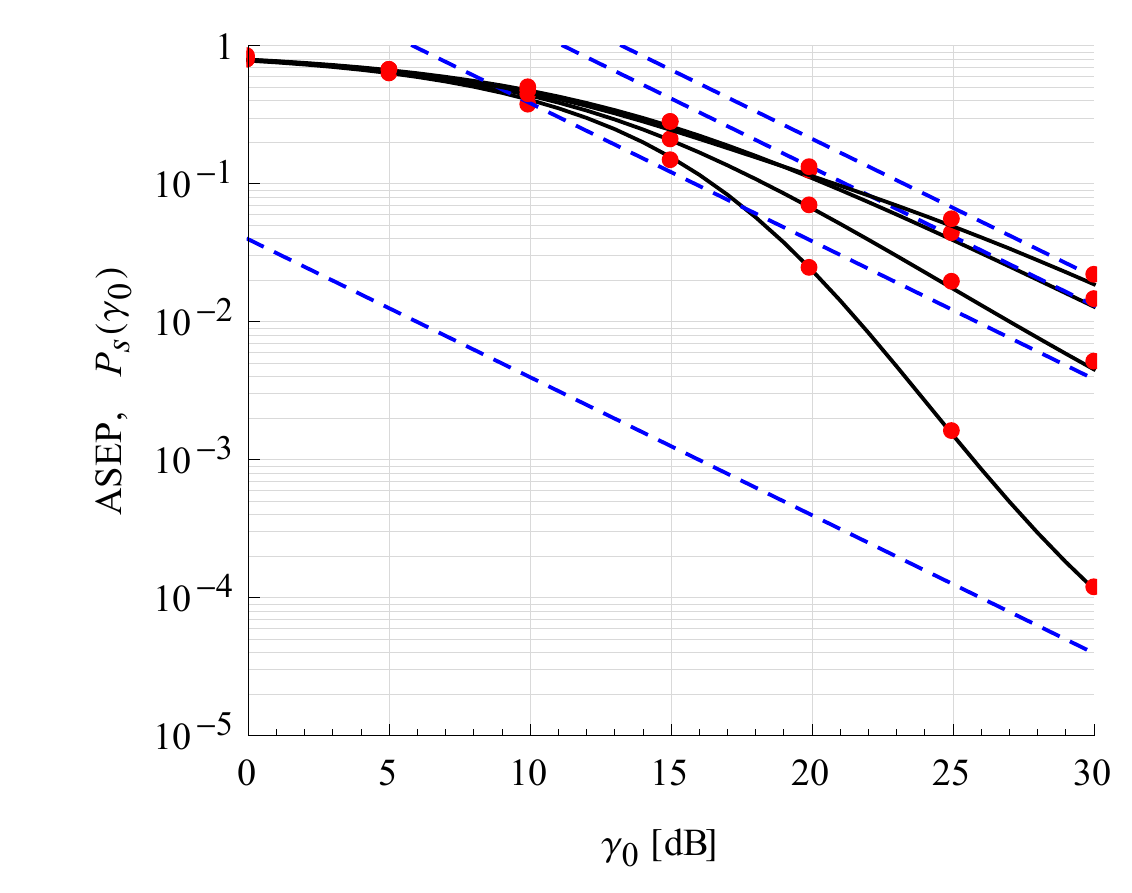}};
            \node[left,near start abs,minimum size=25mm,align = left,anchor=base east] at ($ (b) + (0*2.3,0*2.3*1.3594) $) {\small Rayleigh: ($K=0$)~~\phantom{..}};
            \draw[black,thin,rounded corners = 2pt] ($ (b) + (0*2.3,0*2.3*1.3594) $) -- ($ (b) + (0.1*2.3,0*2.3*1.3594) $) -- ($ (b) + (0.1*2.3,0*2.3*1.3594) + (0.95*2.3,0.95*2.3*1.3594)$);
            \coordinate (b) at ($ (b) + (0*2.3,-0.12*2.3*1.3594) $);
            \node[left,near start abs,minimum size=25mm,align = left,anchor=base east] at ($ (b) + (0*2.3,0*2.3*1.3594) $) {\small Rice: ($K=8, \Gamma=0$)};
            \draw[black,thin,rounded corners = 2pt] ($ (b) + (0*2.3,0*2.3*1.3594) $) -- ($ (b) + (0.1*2.3,0*2.3*1.3594) $) -- ($ (b) + (0.1*2.3,0*2.3*1.3594) + (0.77*2.3,0.77*2.3*1.3594)$);
            \coordinate (b) at ($ (b) + (0*2.3,-0.12*2.3*1.3594) $);
            \node[left,near start abs,minimum size=25mm,align = left,anchor=base east] at ($ (b) + (0*2.3,0*2.3*1.3594) $) {\small ($K=\phantom{1}8, \Gamma=0.5$)~\phantom{...}};
            \draw[black,thin,rounded corners = 2pt] ($ (b) + (0*2.3,0*2.3*1.3594) $) -- ($ (b) + (0.1*2.3,0*2.3*1.3594) $) -- ($ (b) + (0.1*2.3,0*2.3*1.3594) + (1.04*2.3,1.04*2.3*1.3594)$);
            \coordinate (b) at ($ (b) + (0*2.3,-0.12*2.3*1.3594) $);
            \node[left,near start abs,minimum size=25mm,align = left,anchor=base east] at ($ (b) + (0*2.3,0*2.3*1.3594) $) {\small ($K=14, \Gamma=1$)\phantom{.5}~\phantom{...}};
            \draw[black,thin,rounded corners = 2pt] ($ (b) + (0*2.3,0*2.3*1.3594) $) -- ($ (b) + (0.1*2.3,0*2.3*1.3594) $) -- ($ (b) + (0.1*2.3,0*2.3*1.3594) + (1.25*2.3,1.25*2.3*1.3594)$);
      \end{tikzpicture}
        \caption{ }
        %, K=\{0,8,8,14\}, $\Gamma$=\{-,0,0.5,1\}, greška=$10^{-6}$, broj članova=\{2,28,42,77\}}
        \label{ASEP16}
    \end{subfigure}
    \end{tabular} 
    \caption{Exact (solid line) and asymptotic (dashed line) expressions of TWDP ASEP for (a) 2-PSK, (b) 4-PSK, (c) 8-PSK and (d) 16-PSK modulations compared with Monte Carlo simulation results (dots)}
    \label{ASEP}
\end{figure*}
\subsection{Numerical results}
In order to validate the conducted error performance analysis and to justify the proposed parameterization, this section provides graphical interpretation of analytically derived M-ary PSK ASEP and its comparison to results obtained by Monte Carlo simulation. Different modulation orders and TWDP parameters are investigated. 

Fig.~\ref{ASEP2} -~\ref{ASEP16} illustrate the exact (\ref{eq_23}) and the asymptotic (\ref{eq_23'}) ASEP for $2$-PSK, $4$-PSK, $8$-PSK, and $16$-PSK modulations for a set of previously adopted TWDP parameters. ASEP curves, obtained from (\ref{eq_23}) by limiting truncation error to $10^{-6}$, i.e., by employing up to 78 summation terms, are compared with those obtained using Monte Carlo simulations generated with $10^6$ samples. Matching results between the exact and simulated ASEP, as well as between the exact and high-SNR asymptotic ASEP, can be observed for the considered modulation orders and the set of TWDP parameters. Accordingly, derived ASEP expressions can be used to accurately evaluate the error probability of the M-ary PSK receiver for all fading conditions implied by the TWDP model. 

Based on the above, a comparison of error  performance of channels with different fading severities is also performed following Fig.~\ref{ASEP2} -~\ref{ASEP16}. Clearly, the signal in the fading condition characterized with $K = 14$, $\Gamma = 1$ exhibits worse performance compared to the Rayleigh fading channel ($K = 0$), thus representing signal in near hyper-Rayleigh fading conditions. It also can be observed that ASEP in fading conditions described with the same value of $K$ increases with increasing $\Gamma$, indicating that signal performance significantly degrades in channels with $\Gamma= 0.5$ with respect to those in typical Rician channels ($\Gamma = 0$). 

\begin{figure*}[t]
  \centering
    \begin{subfigure}{0.49\linewidth}
    %\centering
      %\includegraphics[trim={0.0cm, 0.2cm, 0.1cm, 0.3cm},clip,width=0.95\linewidth]{PDF.pdf}
      \begin{tikzpicture}[->]
            \coordinate (a) at (1.46,1.075);
            \coordinate (b) at ($ (a) + (1.8*2.3-0.2*2.3,1.0*2.3*1.3594) $);
            \node[anchor=south west,inner sep=0] at (0,0) {\includegraphics[trim={0.0cm, 0.15cm, 0.1cm, 0.3cm},clip,width=0.95\linewidth]{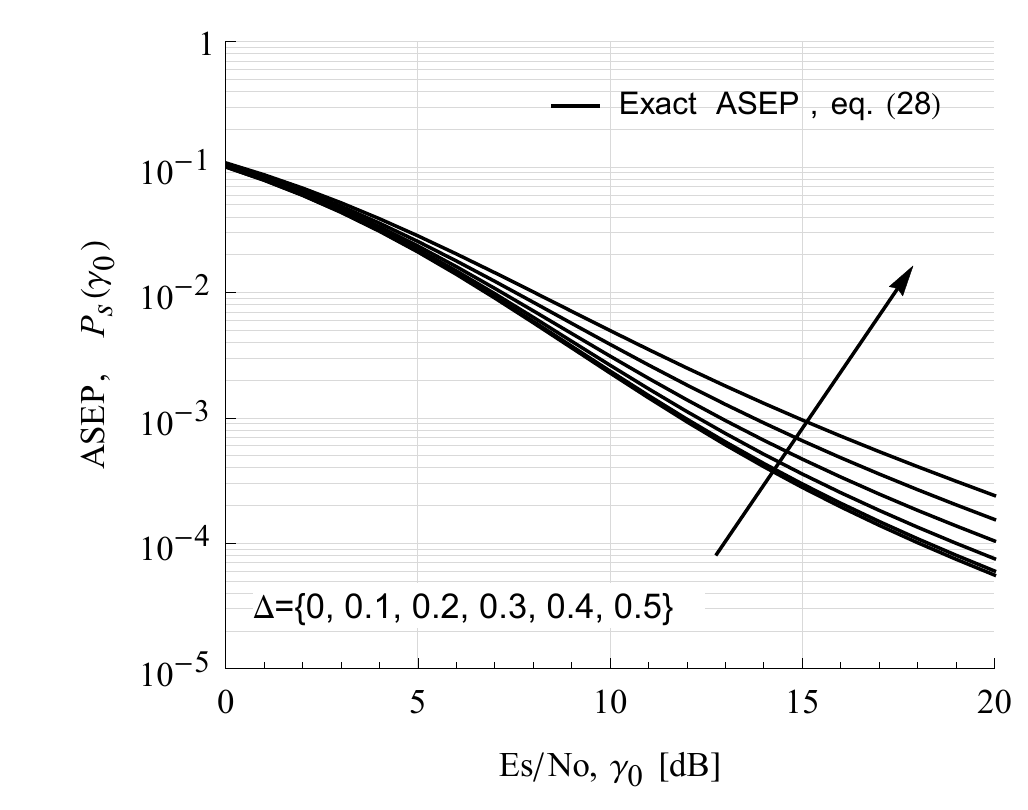}};
      \end{tikzpicture}
      \caption{ }
    \label{Figure6a}
    \end{subfigure}
    \begin{subfigure}{0.49\linewidth}
        \centering
        \begin{tikzpicture}[->]
            \coordinate (a) at (1.50,1.04);
            \coordinate (b) at ($ (a) + (23*0.19846-0.2*2.3,1*1.40) $);
            \node[anchor=south west,inner sep=0] at (0,0) {\includegraphics[trim={0.0cm, 0.1cm, 0.1cm, 0.3cm},clip,width=0.95\linewidth]{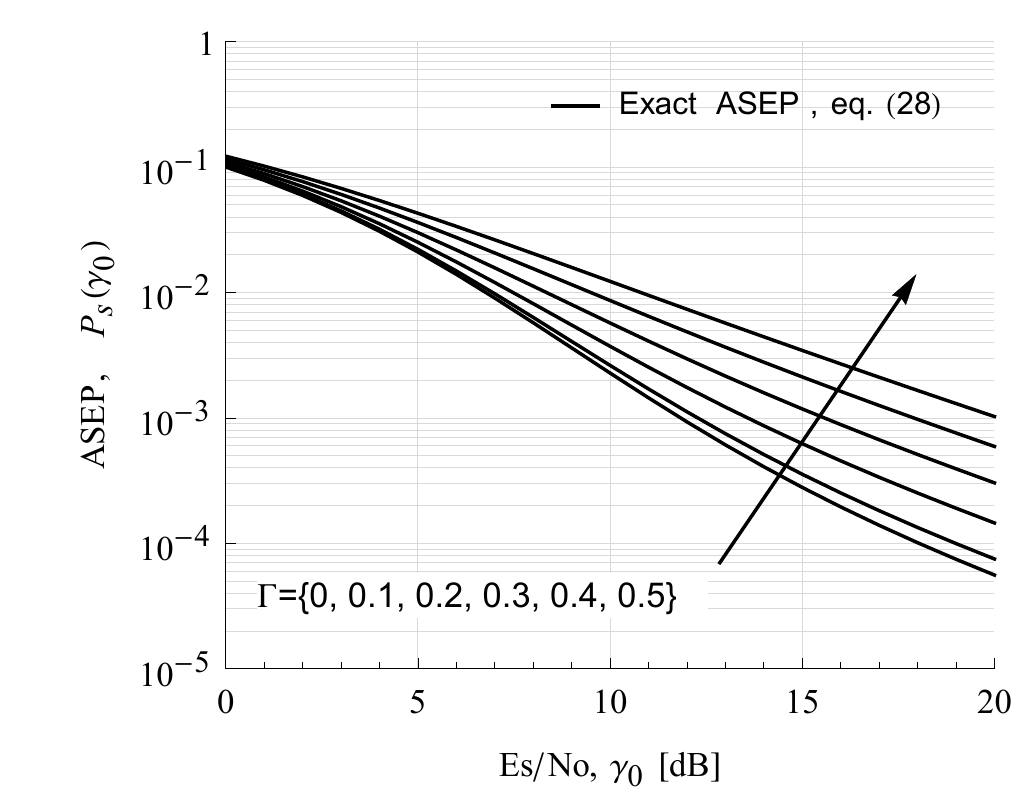}};
      \end{tikzpicture}
        \caption{ }
    \label{Figure6b}
    \end{subfigure}  
  \caption{BPSK ASEP in TWDP channel for $K=6$ and different values of parameter (a) $\Delta$ (b) $\Gamma$ }
  \label{Fig.6}
\end{figure*}
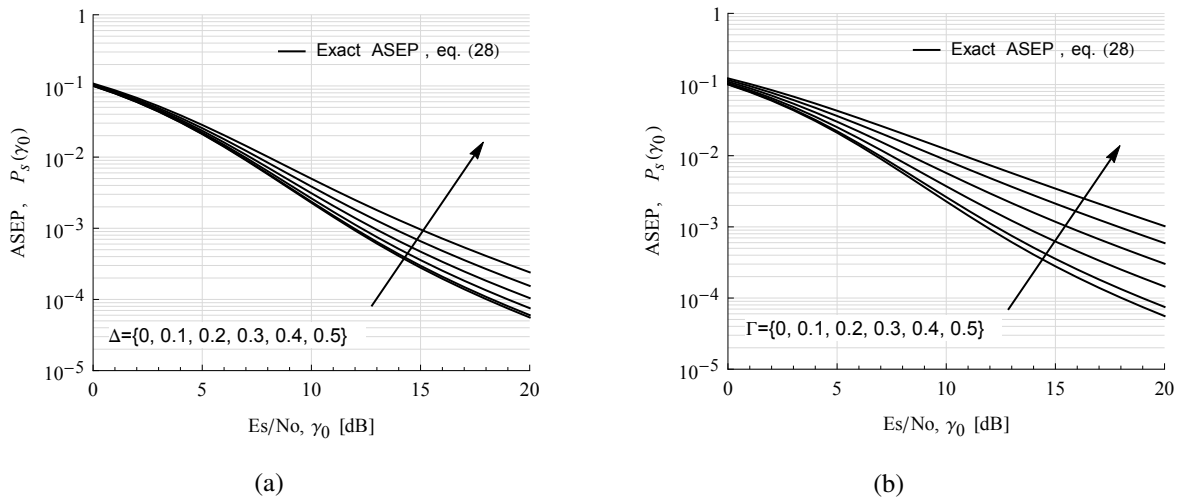

Fig.~\ref{Fig.6} illustrates the effect of proposed parameterization on 2-PSK ASEP curves in TWDP fading channel with $K=6$. Obviously, $\Gamma$-based parameterization solved the problem of densely-spaced ASEP curves observed for the entire range of $\Delta$ between $0$ and $0.5$. 

\section{Conclusion}
\label{sec:V}
This paper proposed  a novel analytical characterization of TWDP fading channels achieved by introducing physically justified TWDP parameterization and exact PDF and CDF expressions, and by deriving the alternative form of the exact SNR MGF expression. Benefits of the proposed parameterization are demonstrated on TWDP PDF and ASEP graphical interpretations. A derived MGF is used for derivation of the exact M-ary PSK ASEP expression, which can be used to accurately evaluate the error performance of M-ary PSK in various fading conditions. 

\section*{Acknowledgment}
The authors would like to thank Prof. Ivo M. Kostić for many valuable discussions and advice.

%\ifCLASSOPTIONcaptionsoff
%  \newpage
%\fi



%\bibliographystyle{IEEEtran}
\bibliographystyle{IEEEtran}
\bibliography{Literatura}

\newcommand{\pomak}{-0mm}
\end{document}